\documentclass[twocolumn]{aastex63}
\usepackage{lineno}

\usepackage{natbib}
\usepackage{amsmath}
\usepackage{url}
\usepackage{graphicx}
\usepackage{color}

\newcommand{\kms}{\,km\,s$^{-1}$}

\def\lsim{\hbox{\rlap{\raise 0.425ex\hbox{$<$}}\lower 0.65ex\hbox{$\sim$}}}
\def\gsim{\hbox{\rlap{\raise 0.425ex\hbox{$>$}}\lower 0.65ex\hbox{$\sim$}}}

\def\arcsec{\hbox{$^{\prime\prime}$}}
\def\arcdeg{\mbox{$^\circ$}}

\newcommand{\hvf}{\text{SN~2020hvf}}

\newcommand{\ISEF}{ISEF International Fellowship}

\shorttitle{\hvf: An Asymmetric SN~Ia with a Surviving Companion}
\shortauthors{Siebert et al.}

\graphicspath{{./}{figures/}}

\begin{document}

\title{An Asymmetric Double-Degenerate Type Ia Supernova Explosion with a Surviving Companion Star}

\correspondingauthor{Matthew~R.~Siebert}
\email{msiebert@ucsc.edu}

\author[0000-0003-2445-3891]{Matthew~R.~Siebert}
\affiliation{Space Telescope Science Institute, Baltimore, MD 21218, USA}

\author[0000-0002-2445-5275]{Ryan J.\ Foley}
\affiliation{Department of Astronomy and Astrophysics, University of California, Santa Cruz, CA 95064, USA}

\author[0000-0002-0632-8897]{Yossef Zenati}
\altaffiliation{\ISEF}
\affil{Physics and Astronomy Department, Johns Hopkins University, Baltimore, MD 21218, USA}

\author[0000-0001-9494-179X]{Georgios Dimitriadis}
\affiliation{School of Physics, Trinity College Dublin, The University of Dublin, Dublin 2, Ireland}

\author{Eva~Schmidt}
\affiliation{Department of Astronomy and Astrophysics, University of California, Santa Cruz, CA 95064, USA}

\author[0000-0001-7823-2627]{Grace~Yang}
\affiliation{The Thacher School, 5025 Thacher Road, Ojai, CA 93023, USA}

\author[0000-0002-5680-4660]{Kyle W. Davis}
\affiliation{Department of Astronomy and Astrophysics, University of California, Santa Cruz, CA 95064, USA}

\author[0000-0002-5748-4558]{Kirsty Taggart}
\affiliation{Department of Astronomy and Astrophysics, University of California, Santa Cruz, CA 95064, USA}

\author[0000-0002-7559-315X]{C\'esar Rojas-Bravo}
\affiliation{Department of Astronomy and Astrophysics, University of California, Santa Cruz, CA 95064, USA}

\begin{abstract}

We present nebular spectroscopy of \hvf{}, a Type Ia supernova (SN~Ia) with an early bump in its light curve. \hvf{} shares many spectroscopic and photometric similarities to the carbon-rich high-luminosity ``03fg-like'' SNe~Ia. At $>$240~days after peak brightness, we detect unambiguous emission from [\ion{Ca}{2}] $\lambda\lambda$7291, 7324 which is rarely observed in normal-SNe~Ia and only seen in peculiar subclasses. \hvf{} displays ``saw-tooth"  emission profiles near 7300~\AA\ that cannot be explained with single symmetric velocity components of [\ion{Fe}{2}], [\ion{Ni}{2}], and [\ion{Ca}{2}], indicating an asymmetric explosion. The broad [\ion{Ca}{2}] emission is best modeled by two velocity components offset by 1,220~km~s$^{-1}$, which could be caused by ejecta associated with each star in the progenitor system, separated by their orbital velocity. For the first time in a SN~Ia, we identify narrow (${\rm FWHM} = 180\pm40$~km~s$^{-1}$) [\ion{Ca}{2}] emission, which we associate with a wind from a surviving, puffed-up companion star.  Few published spectra have sufficient resolution and signal-to-noise ratio necessary to detect similar narrow [\ion{Ca}{2}] emission, however, we have detected similar line profiles in other 03fg-like SNe~Ia. The extremely narrow velocity width of [\ion{Ca}{2}] has only otherwise been observed in SNe~Iax at late times. Since this event likely had a double-degenerate ``super-Chandrasekhar" mass progenitor system, we suggest that a single white dwarf (WD) was fully disrupted and a wind from a surviving companion WD is producing the observed narrow emission. It is unclear if this unique progenitor and explosion scenario can explain the diversity of 03fg-like SNe~Ia, potentially indicating that multiple progenitor channels contribute to this subclass. 

\end{abstract}


\section{Introduction}\label{s:intro}

Thermonuclear white dwarf (WD) explosions are surprisingly diverse considering that a majority, the Type Ia supernovae (SNe~Ia), vary simply enough such that they can be used to measure precise cosmological distances \citep{Phillips93, Riess96}. Despite their important role in measuring the content of the Universe, the exact progenitor system and explosion mechanism of SNe~Ia are not yet well-understood. While it is accepted that SNe~Ia is the product of the thermonuclear explosions of WDs \citep[WDs; see e.g., ][]{Nomoto+84, Livio2000, wang2010, maoz2014observational, Livio_Mazzali+18, Soker19, Jha+19, Ruiter20}, it is still unclear whether the progenitor systems are single- or double-degenerate (SD or DD); i.e., whether the companion is non-degenerate, such as a main-sequence star or red giant\citep[][]{1973ApJ...186.1007W, Mazzali+07}, or a second degenerate WD \citep[][]{IbenTotukov84, Webbink1984, Fink+07, Moll_Woosley13, Holcomb+13_SHedet, Liu+17}. 

After correcting for light-curve shape, color, and host-galaxy properties, SNe~Ia can provide distances with $\sim$8\% uncertainty \citep{Jones19, Scolnic22}. However, there are other WD SNe that do not follow these trends: the low-luminosity, low-ejecta mass SNe~Iax \citep{Foley13:iax}; the low-luminosity, high-ejecta mass SN~2006bt-like SNe \citep{Foley10:06bt}; the high-luminosity, high-ejecta mass ``super-Chandrasekhar'' (SC, alternatively ``03fg-like") SNe~Ia \citep{Howell09}; the low-luminosity, cool SN~2002es-like SNe \citep{Ganeshalingam10}; and the low-luminosity ``Ca-rich'' SNe \citep[see,][]{Perets10:05e, Kasliwal12, WynnV+20b, Zenati+23_carich}.

Photometric and spectroscopic coverage of nearby SNe~Ia has improved to the point that the nearest events are often observed within a few days of the explosion to years after maximum light. Recent attention has been focused on a handful of SNe~Ia that display early excess flux or bumps in their UV/optical light curves relative to a power-law rise. These have been observed in normal-SNe~Ia \citep{Marion+16, Dimitriadis19:18oh_k2, Hosseinzadeh17:17cbv,Miller18, WangQ+23} and a high fraction of peculiar SNe~Ia \citep{Cao15, Jiang17, De19,  Miller20, Jiang+21_SN20hvf, Srivastav23, Dimitriadis23}. Furthermore, the diversity of early excess flux features in their strength, duration, and color likely suggests that multiple mechanisms can lead to their formation \citep{Jiang+18}.

The variety of observed WD explosions is matched by the expansive set of theoretical models of different progenitor systems and explosion mechanisms. In particular, all WD SNe should require a binary system so that a WD can accrete material to the point of He or C ignition, although the companion star could be a main-sequence star, a red giant, a subdwarf, a He star, or another WD. The explosion itself can begin in the center of the WD, slightly off-center, on its surface, or in an accretion stream \citep[e.g,][]{Livio_Mazzali+18, Soker19, Liu_Roepke_Han23_review}. The burning flame can be subsonic (a deflagration), or supersonic (a detonation), and it can transition from a deflagration to a detonation. This delayed-detonation scenario has been used to explain the observed high abundance of intermediate-mass elements (IME) in the SN ejecta since a pure M$_{Ch}$ WD detonation would only synthesize iron group elements \citep{Khokhlov91}.

For some explosion models, the primary WD may survive with delayed radioactive decay-producing super-Eddington winds \citep{Shen_Schwab17}. A non-degenerate star is expected to survive the explosion, but the SN can ablate the star, sweeping up H and/or He that may be observable once the ejecta is optically thin \citep{Mattila+05}. In the DD scenario, the two WDs may violently merge, resulting in a potentially luminous, high-ejecta mass explosion \citep{Pakmor+13, Pakmor_Zenati+21}. Such a system would expel a small amount of material from the secondary WD into the circumstellar medium (CSM), resulting in circumstellar interaction \citep{Raskin13} visible at early times \citep{Dimitriadis21} or as nebular emission at late times \citep{Taubenberger13:10lp}. Alternatively, as gravitational radiation brings the secondary towards the primary, it will begin dynamically transferring its outer non-/semi-degenerate layers, possibly leading to a surface detonation \citep{Guillochon13, Perets_Zenati+19}. This latter scenario could leave the secondary star relatively unscathed but unbound from the system, now traveling through space at its prior orbital velocity \citep[$\sim$1000~km~s$^{-1}$;][]{Shen2021}.

Existing theoretical models have trouble describing all of the characteristic properties of the high-luminosity carbon-rich 03fg-like subclass of SNe~Ia. No formal sub-classification criteria exist for these events, but these SNe often exhibit broad light-curves ($\Delta m_{15} (B) < 1.3$~mag) and high peak luminosities (M$_{B} > -19.5$ mag, \citealt{Howell06, Scalzo10, Yamanaka09}), strong carbon absorption and low absorption velocities at early times \citep{Hicken07, Silverman11:09dc, Chakradhari14, Parrent16, Taubenberger19, Ashall21}, low-ionization nebular spectra \citep{Taubenberger13, Ashall21}, forbidden Calcium emission at late times \citep{Taubenberger19, Dimitriadis21, Dimitriadis23}, and early flux excesses \citep{Jiang17, Chen19, Jiang+21_SN20hvf, Srivastav23}. These SNe also tend to occur in either low-mass host galaxies or remote locations in more massive galaxies potentially indicating a lower metallicity progenitor than those of normal-SNe~Ia \citep{Taubenberger11}. The early spectroscopic properties indicate a dense C/O-rich CSM, but the exact progenitor system is debated. Several studies invoke a DD progenitor whose CSM is produced during a dynamical disruption of one of the WDs \citep{Raskin13, Taubenberger13, Dimitriadis21, Dimitriadis23, Srivastav23}, while others suggest instead these SNe are produced from the ejecta interacting with the envelope of an asymptotic giant branch (AGB) star (the ``core-degenerate" scenario \citealt{Kashi_Soker11, Hsiao20, Ashall21}). Each of these scenarios has drawbacks. In the former case, one would expect a larger degree of asymmetry in the ejecta distribution which has not been observed with continuum polarization measurements \citep{Cikota19}. In the latter case, no signs of interaction, such as narrow emission features, have been observed at late times. 

A common theme for nearly all WD SNe is their symmetry. Polarization measurements of WD SNe and the shape of SN remnants indicate that WD explosions are usually nearly spherical \citep{Leonard06, Wang_Wheeler08, Bulla_Sim_Kromer15, Bulla19}. Nevertheless, there are indications in the ejecta velocity distribution of SNe~Ia \citep{Foley11:vgrad}, and the correlation between photospheric velocity, velocity gradient, velocity shifts in late-time spectra, and polarization that SNe~Ia are mildly asymmetric \citep{Maeda10:asym}. In particular, velocity shifts in late-time, nebular spectra when the ejecta is optically thin are a window to the line-of-sight (LOS) kinematics of the ejecta. In addition to offsets indicative of an off-center explosion, some SNe~Ia (particularly those with low luminosity) have apparent multiple-component nebular features \citep[although the interpretation has been disputed since this is not seen in all features simultaneously]{Dong15, Tucker20:nebular}. Multiple kinematic components would indicate a clearly asymmetric explosion.

Here we present optical observations of the \hvf{}. \hvf{} was discovered on 21 April 2020 \citep{Tonry+20_hvf} by the Asteroid Terrestrial-impact Last Alert System \citep[ATLAS;][]{Tonry+18_ATLAS}. Its host galaxy, NGC~3643, has a redshift of $z = 0.005811$ \citep{vanDriel16} and a distance of $29.4\pm 2.1$~Mpc (measured using the NASA/IPAC Extragalactic Database, NED\footnote{http://ned.ipac.caltech.edu/}). It was promptly classified as a young SN~Ia with strong C features \citep{Burke20_20hvf}. High-cadence photometry starting immediately after discovery revealed a luminous, day-long flash, which was interpreted as interaction with circumstellar material \citep[CSM;][]{Jiang+21_SN20hvf}. The early light-curve evolution and photospheric spectra of \hvf{} are similar to 03fg-like SNe~Ia \citep{Jiang+21_SN20hvf, Srivastav23}.

The data presented here extend the spectroscopic coverage of \hvf{} to 332~days after peak brightness. We present several late-time, nebular spectra, which are qualitatively similar to that of 03fg-like SNe~Ia. Examining the nebular line profiles in detail, we find three distinct kinematic components, indicating an asymmetric explosion and a surviving WD.

We present our observations in Section~\ref{s:obs}.  In Section~\ref{s:anal}, we examine our full data set. We discuss our results in Section~\ref{s:disc} and conclude in Section~\ref{s:conc}.

\section{Observations \& Data Reduction}\label{s:obs}

\begin{figure}
\begin{center}
    \includegraphics[width=3.2in]{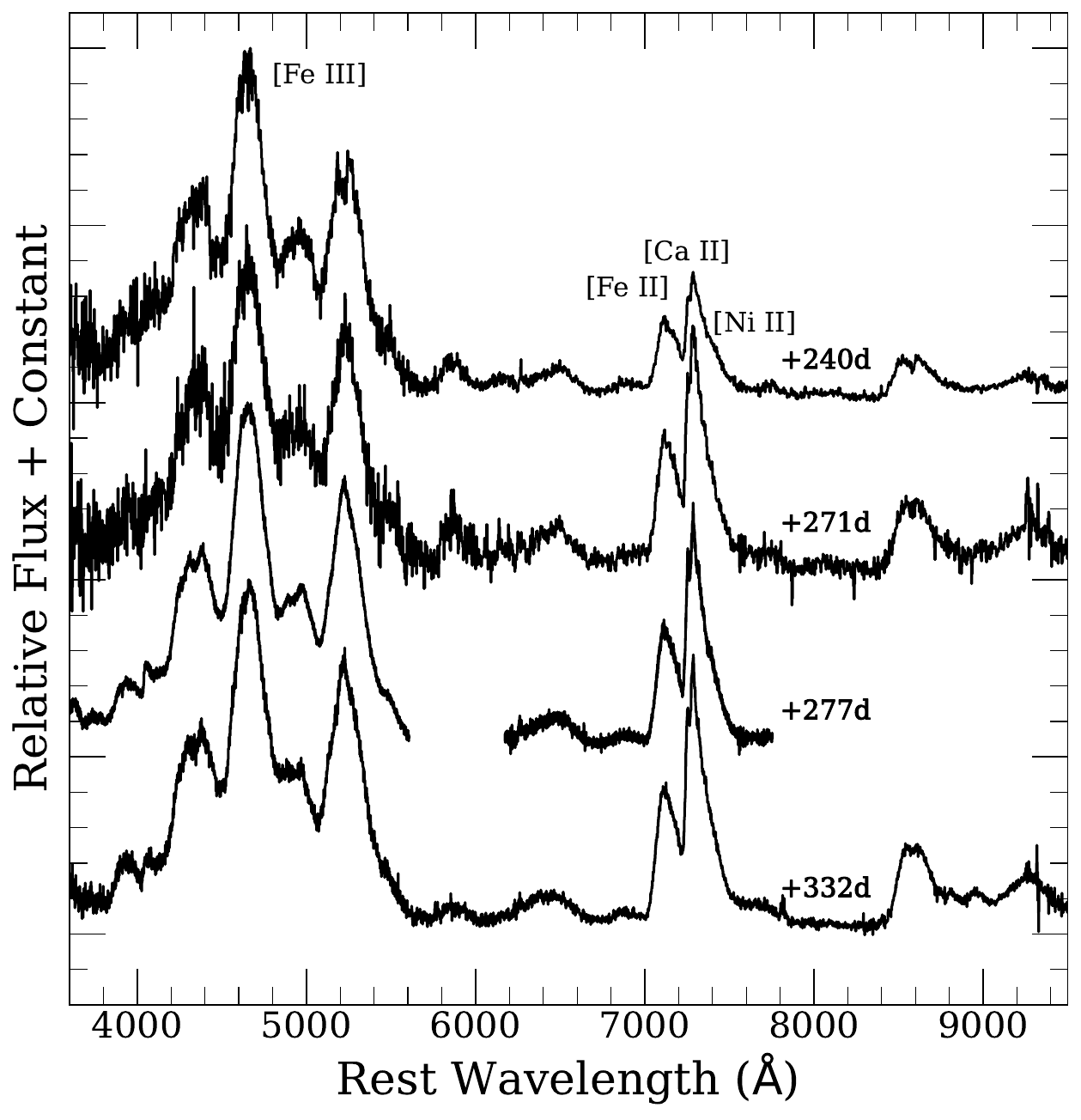}
\caption{Nebular spectra of \hvf{} ranging from +240-332 days after peak brightness. Data were acquired from the Kast spectrograph on the Lick Shane telescope and the LRIS spectrograph on the Keck I telescope. }\label{f:neb_spec}
\end{center}
\end{figure}

We obtained a total of 4 optical spectra of \hvf{} that range from 240 to 332 days after maximum light. Two spectra were acquired with the Kast spectrograph \citep{Miller93} on the Lick Shane telescope and two spectra were acquired with the LRIS with the Low-Resolution Imaging Spectrometer \citep[LRIS;][]{Oke95}, mounted on the 10-meter Keck~I telescope at the W.\ M.\ Keck Observatory. The Kast spectra were obtained on 2021 Jan 6 and 2021 Feb 6 UT and were both observed with the 452/3306 grism and 300/7500 grating. These observations used the 2\arcsec-wide slit and d57 dichroic and cover a wavelength range of 3,255 -- 10,893~\AA. The LRIS observations were obtained with two different setups. On 2021 Feb 12 UT,  we observed a  low-resolution setting (1800 and 1430~s blue channel exposures with the B600/4000 grism) and a high-resolution red channel setting (two 825~s exposures with the R1200/7500 grating, with a pixel scale of 0.4~\AA/pixel). The blue channel and red channel cover spectra cover 3,141 -- 5,640~\AA and  6,210 -- 7,800~\AA, respectively. On 2021 Apr 8 UT we observed an identical blue channel setup, but obtained two 525~s red channel exposures with the R400/8500 grating, with pixel scales of 0.63 and 1.16~\AA/pixel, respectively). This low-resolution spectrum covers 3,162 -- 10,147~\AA. We used the 1.0\arcsec-wide slit and the D560 dichroic for all observations and oriented the slit to include the host-galaxy nucleus. The atmospheric dispersion corrector unit was deployed. Each of these spectra includes H$\alpha$, \ion{He}{1} $\lambda$6678, [\ion{O}{1}] $\lambda\lambda$6300, 6364 and the 7300~\AA\ line complex, the primary focus of our current analysis.  All data were reduced using the UCSC Spectral Pipeline \footnote{\url{https://github.com/msiebert1/UCSC_spectral_pipeline}} which makes use of standard \textsc{iraf}\footnote{IRAF is distributed by the National Optical Astronomy Observatory, which is operated by the Association of Universities for Research in Astronomy (AURA) under a cooperative agreement with the National Science Foundation.} and python routines for bias/overscan corrections, flat fielding, flux calibration, and telluric lines removal, using spectro-photometric standard star spectra, obtained the same night \citep{Silverman12:bsnip}. We present these spectra in \autoref{f:neb_spec}. 

\section{Analysis}\label{s:anal}

\subsection{Distance and Luminosity}

NGC~3643 is close enough that a redshift-derived distance could be significantly biased. However, there are no redshift-independent distance measurements for NGC~3643 in the literature. \citet{Jiang+21_SN20hvf} used the NED distance calculator, correcting for the Virgo Supercluster, Great Attractor, and Shapley Supercluster to find a distance modulus of $\mu = 32.45 \pm 0.15$~mag, corresponding to $D = 30.9$~Mpc. Literature values for NGC~3643 under different peculiar velocities and Hubble constant assumptions result in similar results of $\mu = 32.1$~mag \citep{Kourkchi17, Leroy19}. Using the Numerical Acion Methods velocity field reconstruction \citep{Shaya17} for the Cosmicflows-3 catalog \citep{Kourkchi20} through the Extragalactic Distance Database \citep{Tully09} and assuming $H_{0} = 73.04$~km~s$^{-1}$~Mpc$^{-1}$ \citep{Riess16, Riess18:gaia}, we find $\mu = 32.38 \pm 0.13$~mag ($D = 29.96$~Mpc), which we consider the best value for the distance to \hvf{}.  Using this distance, we find $M_g= -19.91$ mag which is consistent with \citet{Jiang+21_SN20hvf}.

\subsection{Spectroscopic comparisons}

\begin{figure*}
\begin{center}
    \includegraphics[width=6.8in]{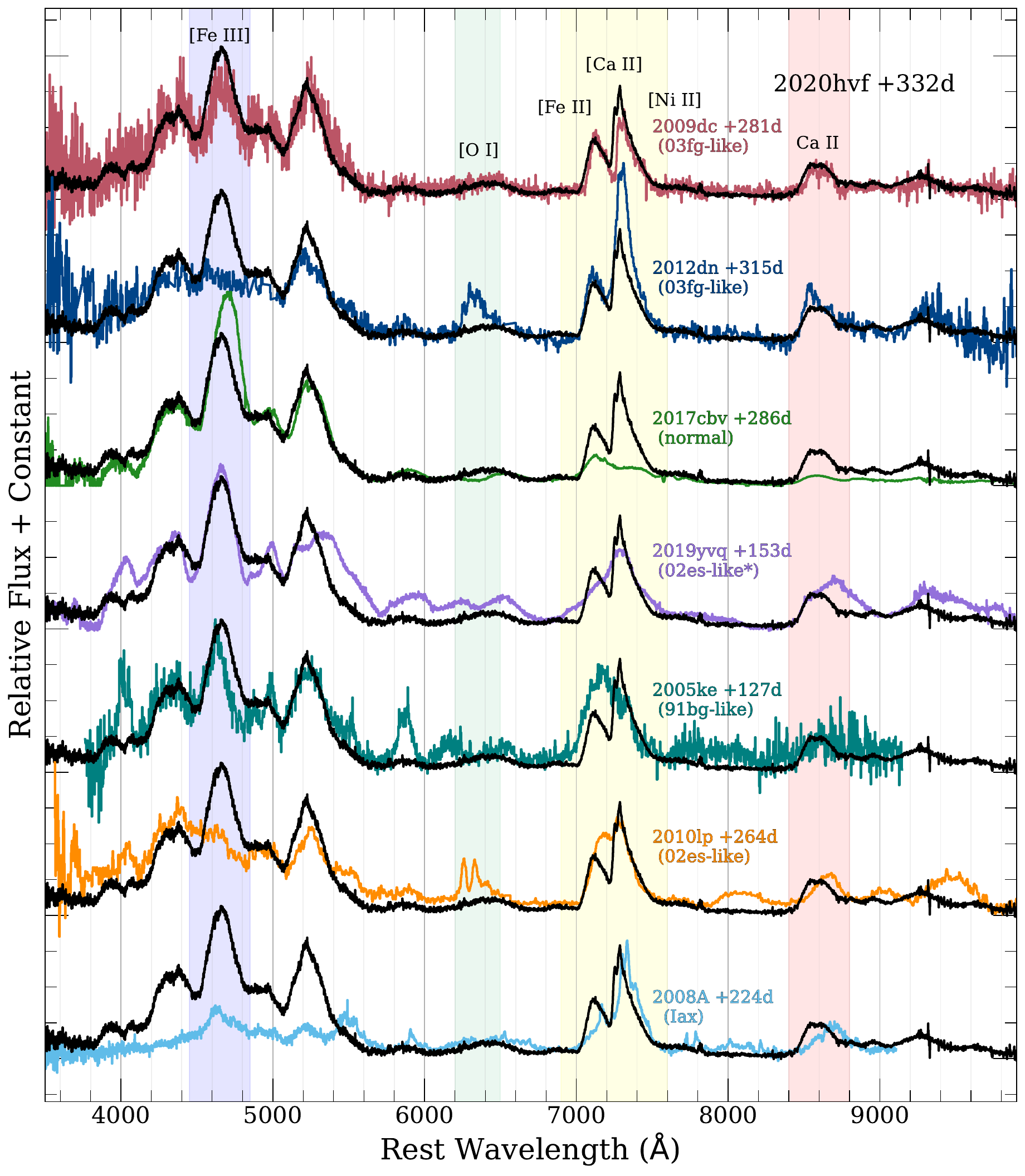}
\caption{Optical spectrum of \hvf{} (black curves) at +332~days after peak brightness compared to those of other SNe~Ia at similar phases. From top to bottom, we compare scaled nebular spectra of in order decreasing peak $B$-band brightness: SN~2009dc (red) and SN~2012dn (blue) which are both 03fg-like SNe~Ia; SN~2017cbv which also had an early-time flux excess (green);  SN~2019yvq (purple), a peculiar SN~Ia with strong [\ion{Ca}{2}] (*classified as ``transitional 02es-like" by \citealt{Burke21}); SN~2005ke (teal), a 91bg-like SN~Ia; SN~2010lp (orange), a peculiar SN~2002es-like SN which had nebular [\ion{O}{1}] emission (orange); and SN~2008A (light blue), a SN Iax with narrow nebular features. Where relevant, we have clipped emission lines from the host galaxy for better visualization. Several spectral regions are highlighted: [\ion{Fe}{3}] $\lambda 4701$ (blue); [\ion{O}{1}] $\lambda\lambda$6300, 6364 (green); the feature at 7300~\AA\ complex which includes possible contributions from [\ion{Fe}{2}] $\lambda 7155$, [\ion{Ni}{2}] $\lambda 7378$, and [\ion{Ca}{2}] $\lambda \lambda 7291$, 7324 (yellow); and the \ion{Ca}{2} NIR triplet (red). }\label{f:spec_comp}
\end{center}
\end{figure*}

The nebular features of \hvf{} show interesting similarities to other 03fg-like and peculiar SNe~Ia. \autoref{f:spec_comp} displays our +332d LRIS spectrum of \hvf{} (black) in comparison to other thermonuclear SNe in order of luminosity. From top to bottom we show the 03fg-like SN~Ia 2009dc ($M_{B, {\rm ~peak}} = -20.3$~mag; the 03fg-like SN~Ia 2012dn ($M_{B, {\rm ~peak}} = -19.28$~mag; \citealt{Taubenberger19})); \citealt{Silverman11:09dc}); SN~2017cbv ($M_{B, {\rm ~peak}} = -19.25$~mag; \citealt{Hosseinzadeh17:17cbv, Sand18}), a normal SN~Ia with an early light curve flux excess; SN~2019yvq, a peculiar SN~Ia with strong [\ion{Ca}{2}] emission in its nebular spectrum ($M_{g, {\rm ~peak}} = -18.5$~mag; \citealt{Miller20,Siebert20b}); the 91bg-like peculiar SN~Ia 2005ke ($M_{V, {\rm ~peak}} = -17.9$~mag; \citealt{Silverman12:bsnip}); SN~2010lp ($M_{B, {\rm ~peak}} = -17.7$~mag; \citealt{Kromer13}; Pignata et al., in preparation), a 2002es-like SN Ia that had strong, narrow [\ion{O}{1}] emission in its nebular spectrum; and SN~2008A ($M_{V, {\rm ~peak}} = -18.2$~mag; \citealt{Foley16:iax}), a SN Iax with narrow nebular features. The 03fg-like  SNe~Ia in this sample, SN~2009dc and SN~2012dn, both have low [\ion{Fe}{3}]/[\ion{Fe}{2}] ratios (measured from the peak intensity of [\ion{Fe}{3}] $\lambda 4701$ to [\ion{Fe}{2}] $\lambda 7155$) relative to normal SNe~Ia. The weak or lack of [\ion{Fe}{3}] emission is a well-documented feature of this subclass and has been attributed to the low-ionization state that results from the high-central densities of these explosions \citep{Taubenberger19, Ashall21}. The low-ionization in the low-luminosity events SN~2005ke, SN~2010lp, and SN~2008A is likely the result of low overall ejecta temperatures, and despite these events likely having different explosion scenarios, their nebular features are actually quite similar to 03fg-like SNe~Ia. \hvf{} has a clear component of [\ion{Fe}{2}] emission and its [\ion{Fe}{3}]/[\ion{Fe}{2}] ratio is most similar to SN~2009dc ([\ion{Fe}{3}]/[\ion{Fe}{2}] $=$ 2.4 and 1.7, respectively). The width of [\ion{Fe}{3}] in \hvf{} (FWHM$=10,300$\kms) is most similar to that of SN~2019yvq at +153d (FWHM$=10,500$\kms). 

All of the SNe presented in \autoref{f:spec_comp} (aside from SN~2017cbv) likely have contributions from Calcium in their nebular spectra. In particular, the peak of the 7300 \AA\ emission feature in each of these SNe occurs near [\ion{Ca}{2}] $\lambda\lambda$7291, 7324, and strong components of [\ion{Ca}{2}] are easily seen in \hvf{}, SN~2012dn, SN~2019yvq, SN~2010lp, and SN~2008A. Furthermore, these five SNe also show prominent Ca II NIR triplet emissions.

\begin{figure}
\begin{center}
    \includegraphics[width=2.55in]{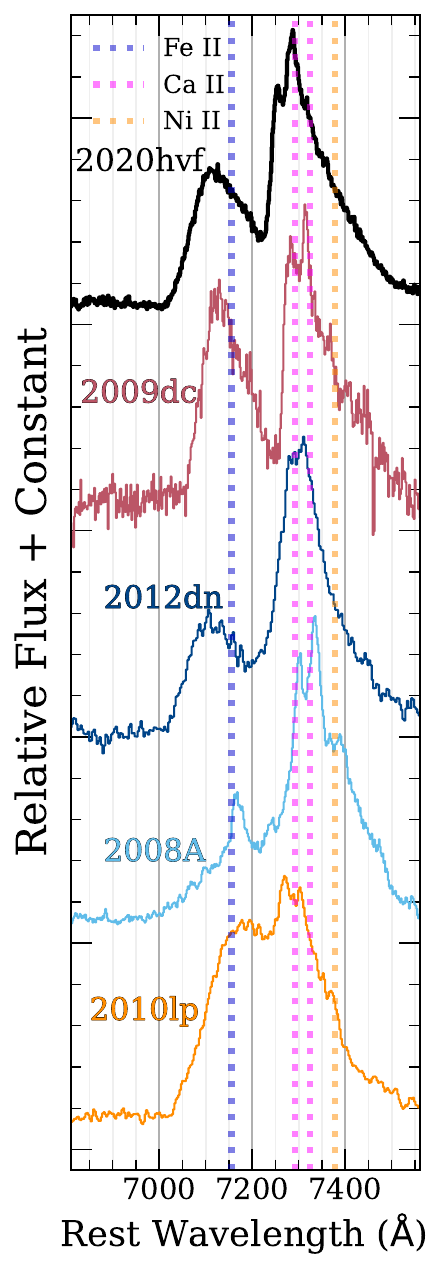}
\caption{The 7300\AA\ emission feature of \hvf{} in comparison to the nebular spectra other 03fg-like and peculiar SNe~Ia from \autoref{f:spec_comp}. From top to bottom the SNe displayed are : \hvf{} (03fg-like, black curve), SN~2009dc (03fg-like, red curve), SN~2012dn (03fg-like, blue curve), SN~2008a (SN~Iax, light blue curve), and SN~2010lp (``02es-like", orange curve).}\label{f:spec_comp_7300}
\end{center}
\end{figure}

We investigate the [\ion{Ca}{2}] emission morphology in the 7300 \AA\ feature in more detail in \autoref{f:spec_comp_7300}. Here we compare \hvf{} to a selection of 03fg-like SNe~Ia (SN~2009dc and SN~2012dn), SN~2010lp, and SN~2008A. The strong lines of [\ion{Fe}{2}], [\ion{Ni}{2}], and [\ion{Ca}{2}], are shown as dashed-blue, orange, and pink vertical lines respectively. It is clear that several of these SNe require components of [\ion{Ca}{2}] emission that are narrower than that of the more isolated [\ion{Fe}{2}] emission. Aside from SN~2008A, the [\ion{Ca}{2}] emission appears blueshifted in each of these SNe, and its strength relative to [\ion{Fe}{2}] varies significantly. This could be a result of different strength contributions from [\ion{Ni}{2}] which could be overlapping with the contribution from [\ion{Ca}{2}]. Additionally, the shape of the [\ion{Fe}{2}] emission in \hvf{} and SN~2009dc does not look Gaussian. Other studies have commented on the characteristic ``sharp" features in 03fg-like SNe~Ia \citep{Chen19, Taubenberger19}, but so far there has not been an effort to characterize these features in detail. \citet{Taubenberger19} suggested that the steeper blue edge and shallow-sloped red edge of [\ion{Ca}{2}] in SN~2012dn could arise from ejecta with an emissivity distribution that is peaked toward the near side of the ejecta. Or alternatively, there was significant dust formation causing the receding ejecta to be more strongly extinguished. In section \ref{ss:fits} we explore the possibility of these features arising from the presence of multiple velocity components of [\ion{Ca}{2}].

\begin{figure}
\begin{center}
    \includegraphics[width=3.2in]{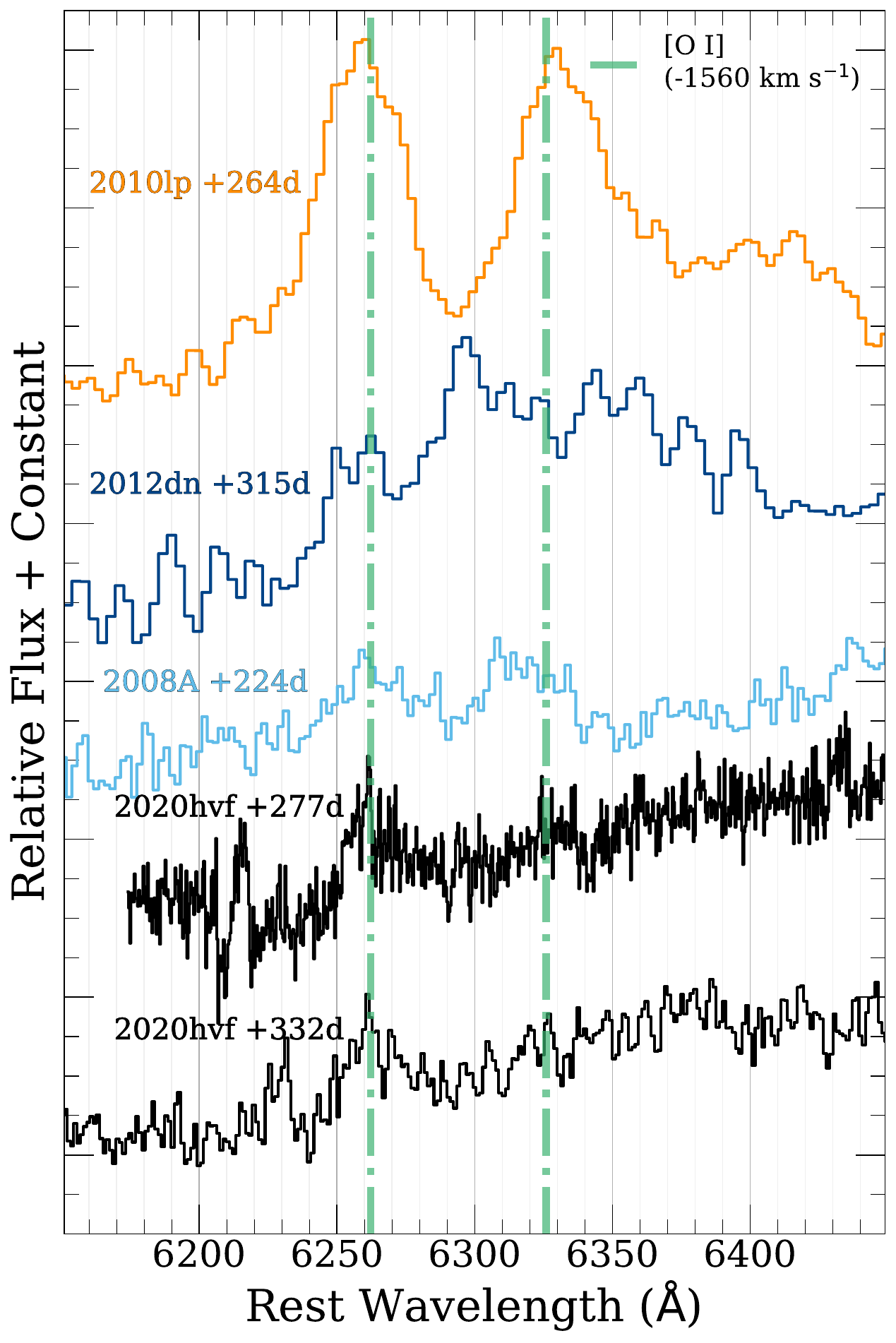}
\caption{Comparison of the nebular LRIS spectra of \hvf{} (black) with other SNe~Ia that may have [\ion{O}{1}] emission. Starting from the top, the other SNe are the ``02es-like" SN~2010lp, the 03fg-like SN~Ia 2012dn, and the SN~Iax 2008A. The dashed green lines show [\ion{O}{1}] $\lambda\lambda$6300, 6364, blueshifted by $1,560$~km~s$^{-1}$. SN~2010lp is the only SN with clearly detected [\ion{O}{1}], however, potential weak emission in \hvf{} and SN~2008A could indicate similar explosion scenarios. }\label{f:OI}
\end{center}
\end{figure}

Given that multiple comparison spectra in \autoref{f:spec_comp} show broad or narrow [\ion{O}{1}] $\lambda\lambda$6300, 6364 emissions, we investigate this emission further and compare to \hvf{} in \autoref{f:OI}. We compare our nebular LRIS spectra of \hvf{} (black curves) to the nebular spectra of the ``02es-like" SN~2010lp (orange), the 03fg-like SN~2012dn (blue), and the SN~Iax 2008A (light blue). The dashed-green lines show [\ion{O}{1}] at $1{,}560$~km~s$^{-1}$, the velocity of the bluest peak of [\ion{Ca}{2}] in the 7300\AA\ emission feature. SN~2010lp is the only SN~Ia with a clear and unambiguous detection of [\ion{O}{1}] \citep{Taubenberger13}. Similar emission was also seen in the ``02es-like" SN~Ia iPTF14atg \citep{Kromer+16}, which also had an early flux excess in its light curve. SN~2012dn appears to have broad (FWHM$\sim5500$~km~s$^{-1}$) [\ion{O}{1}] emission without a significant blueshift \citep{Taubenberger19}. Broad [\ion{O}{1}] emission was also observed in SN~2021zny \citep{Dimitriadis23}, another 03fg-like SN~Ia with an early flux excess. For both \hvf{} and SN~2008A, we identify potential weak [\ion{O}{1}] emission. While the presence of this feature is hard to constrain, we note that for \hvf{}, the [\ion{O}{1}] $\lambda$6300 line is at $-1,800$~km~s$^{-1}$ which is similar to the velocity measured from the bluest peak of [\ion{Ca}{2}]. While the presence of [\ion{O}{1}] is difficult to reconcile with a SC mass explosion, this could indicate further similarity of \hvf{} with the progenitor systems of SNe~Iax like SN~2008A.

\subsection{Ejecta Kinematics from Forbidden Emission Lines}\label{ss:fits}

The emission lines at late times provide important information about the ejecta's velocity profile along the line of sight. Because the ejecta is optically thin at these times, one can use the line profile to determine bulk offsets in the ejecta and aspherical kinematic profiles.

Following the method of \citet{Siebert20b} and \citet{Maguire18}, we attempt to fit the line complex centered at $\sim$7300~\AA\ of our highest resolution spectrum (+277d) with a combination of [\ion{Fe}{2}], [\ion{Ni}{2}] and [\ion{Ca}{2}] lines. There are several individual transitions from these ions in this wavelength regime, and our method properly accounts for their contribution to a blended profile. Our procedure fits a linear continuum across the line complex, assumes that the line profiles are Gaussian, and fits for the velocity offset, velocity width, and strength of the lines (where the relative strength of each line is fixed from the atomic physics). The best-fitting emission line parameters are determined in a two-stage process. First, an initial chi-squared minimization is performed. Then, the parameters determined by this fit are used as input for an MCMC simulation which maximizes the log-likelihood function. Each fit is performed with 50 walkers over 1000 iterations. For priors, we require amplitudes and velocity widths to be greater than 0, and an absolute velocity offset less than $5000$\kms. Best fitting parameters and uncertainties are determined from the walker parameter distributions at the end of the simulation. Given the complex narrow line morphologies in \hvf{}, we have also improved the fitting method so that a Gaussian line spread function (measured from the night sky emission lines in each optical spectrum) is first convolved with each emission line profile. This puts an effective limit on the minimum velocity width we can obtain from each spectrum using this fitting method.

\begin{figure}
\begin{center}
    \includegraphics[width=3.2in]{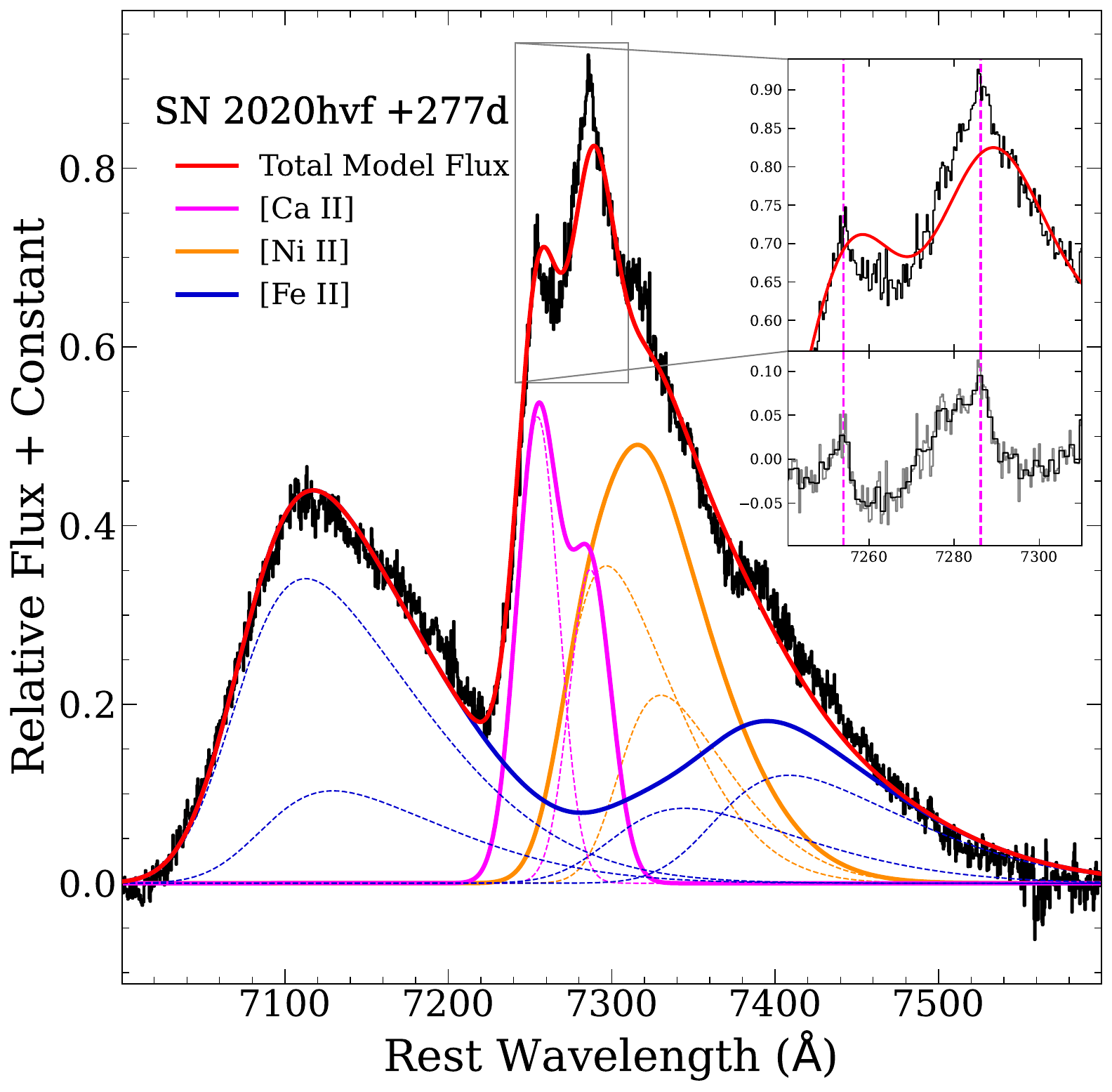}
\caption{Continuum-subtracted 277-d spectrum of \hvf{} covering the Fe/Ni/Ca line complex (black). A simple single-component Fe/Ni/Ca fit is illustrated by the solid and dashed colored curves. Dashed lines show contributions from individual emission lines, and solid lines, show the sum of the emission from each element. The red solid curve represents the entire emission model. A single Gaussian component of [\ion{Ca}{2}] is unable to reproduce the sharp emission features.}\label{f:neb_fit_bad}
\end{center}
\end{figure}

It is obvious that [\ion{Fe}{2}] and [\ion{Ni}{2}] are insufficient to fit the entire complex. There is strong emission at $\sim$7280~\AA\ that cannot be fit by either line without unphysical velocity offsets but is consistent with [\ion{Ca}{2}] emission. This line has been identified in several other peculiar SNe~Ia, and peaks in the line complex corresponding to the wavelength difference for the doublet make this identification particularly robust for \hvf{}.

\begin{figure*}
\begin{center}
    \includegraphics[width=3.3in]{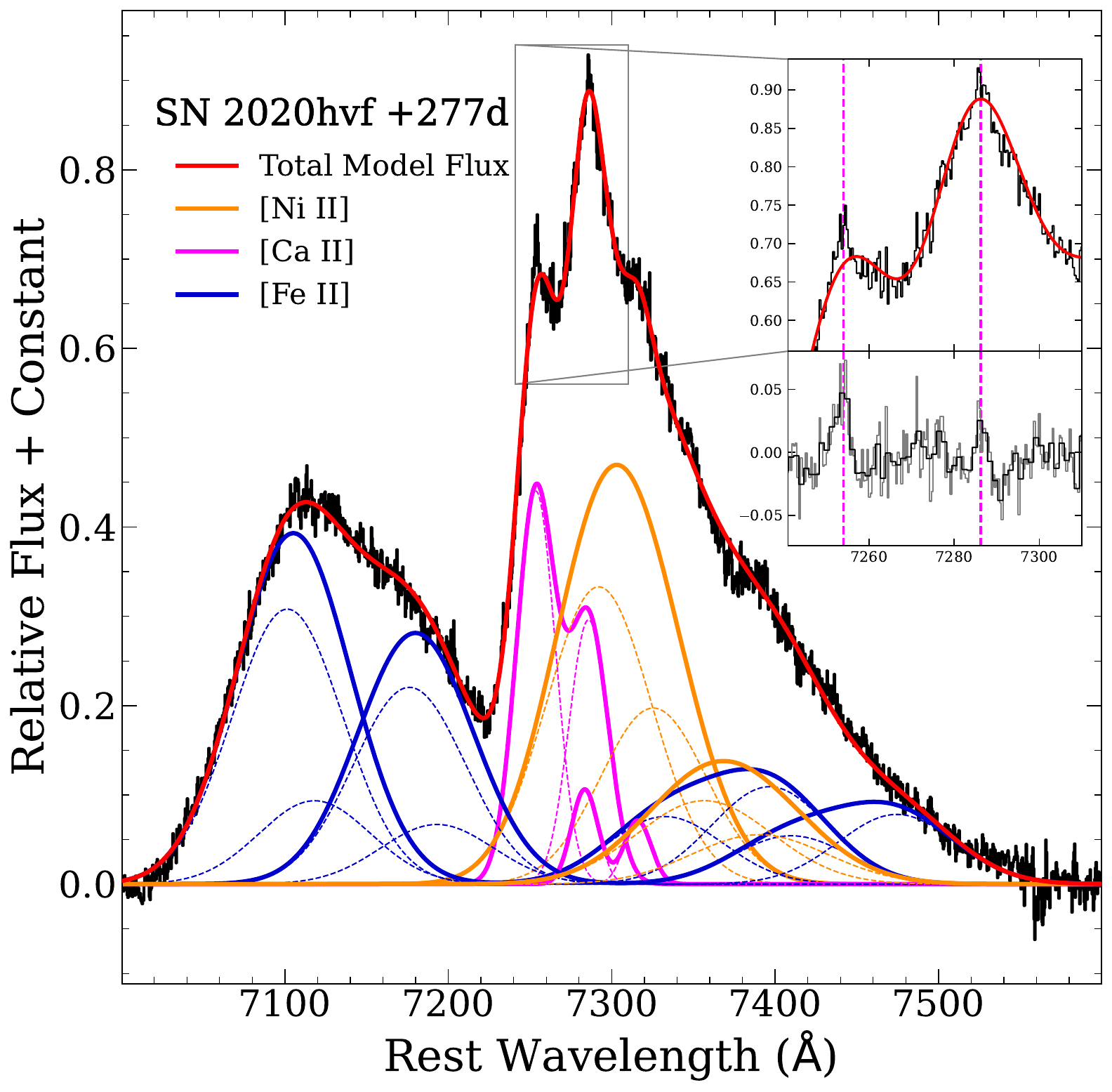}
    \includegraphics[width=3.3in]{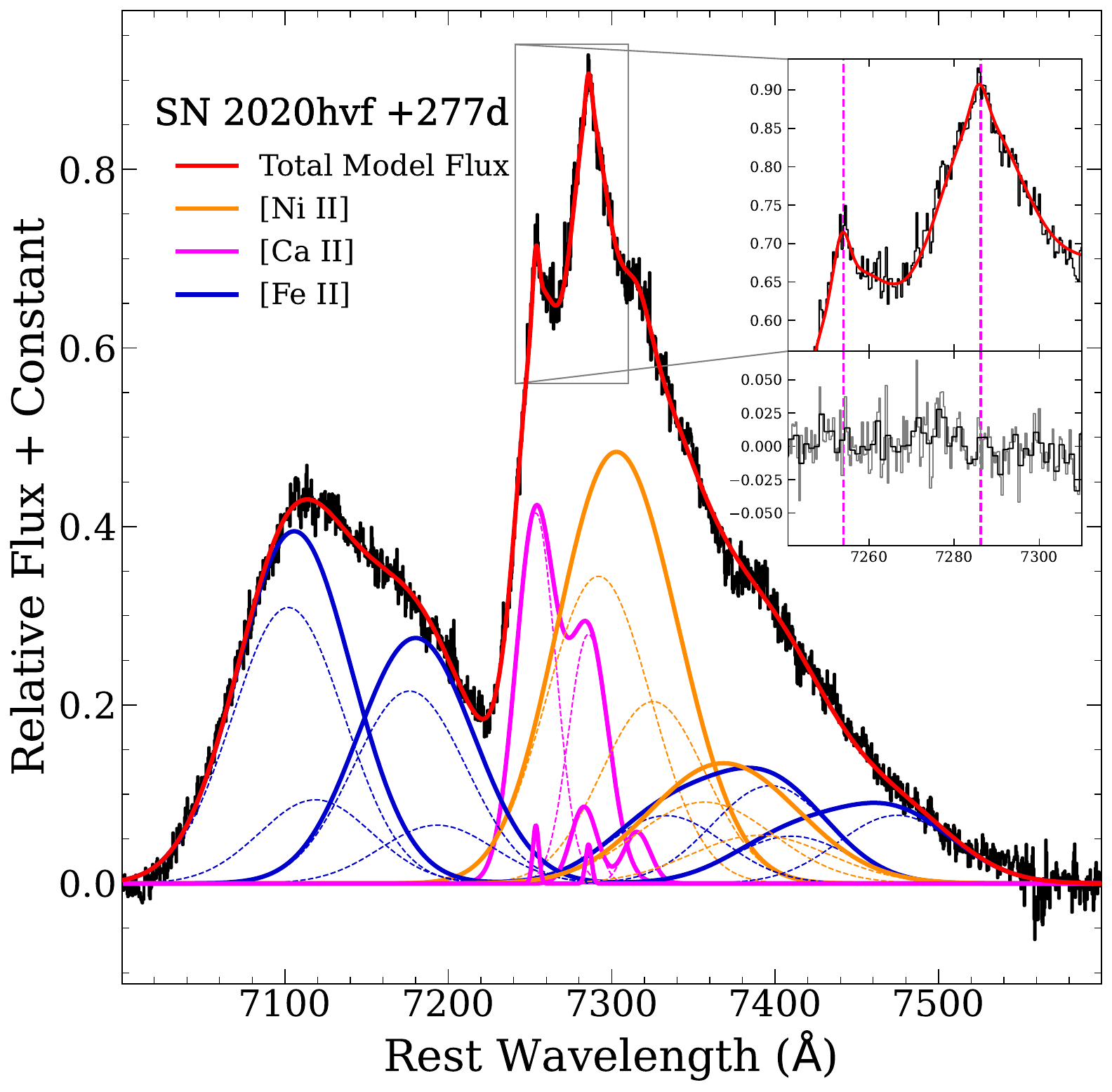}
\caption{\textit{left}: Continuum-subtracted 277-d spectrum of \hvf{} covering the Fe/Ni/Ca line complex. The left panel shows the spectrum in black, as well as a multi-component Gaussian fit (see text) for [\ion{Ca}{2}] (magenta), [\ion{Fe}{2}] (blue), and [\ion{Ni}{2}] (orange), with the combined fit shown in red.  The inset shows the region around the bluer [\ion{Ca}{2}] emission. The bottom panel of the inset shows the residuals to the model with narrow [\ion{Ca}{2}] lines marked. \textit{right}: Same as left plot, but with a third, narrow [\ion{Ca}{2}] component added to the model. The best fitting parameters and uncertainties for each model are detailed in Table \ref{t:neb_params}.}\label{f:neb_fit}
\end{center}
\end{figure*}

\begin{figure}
\begin{center}
    \includegraphics[width=3.3in]{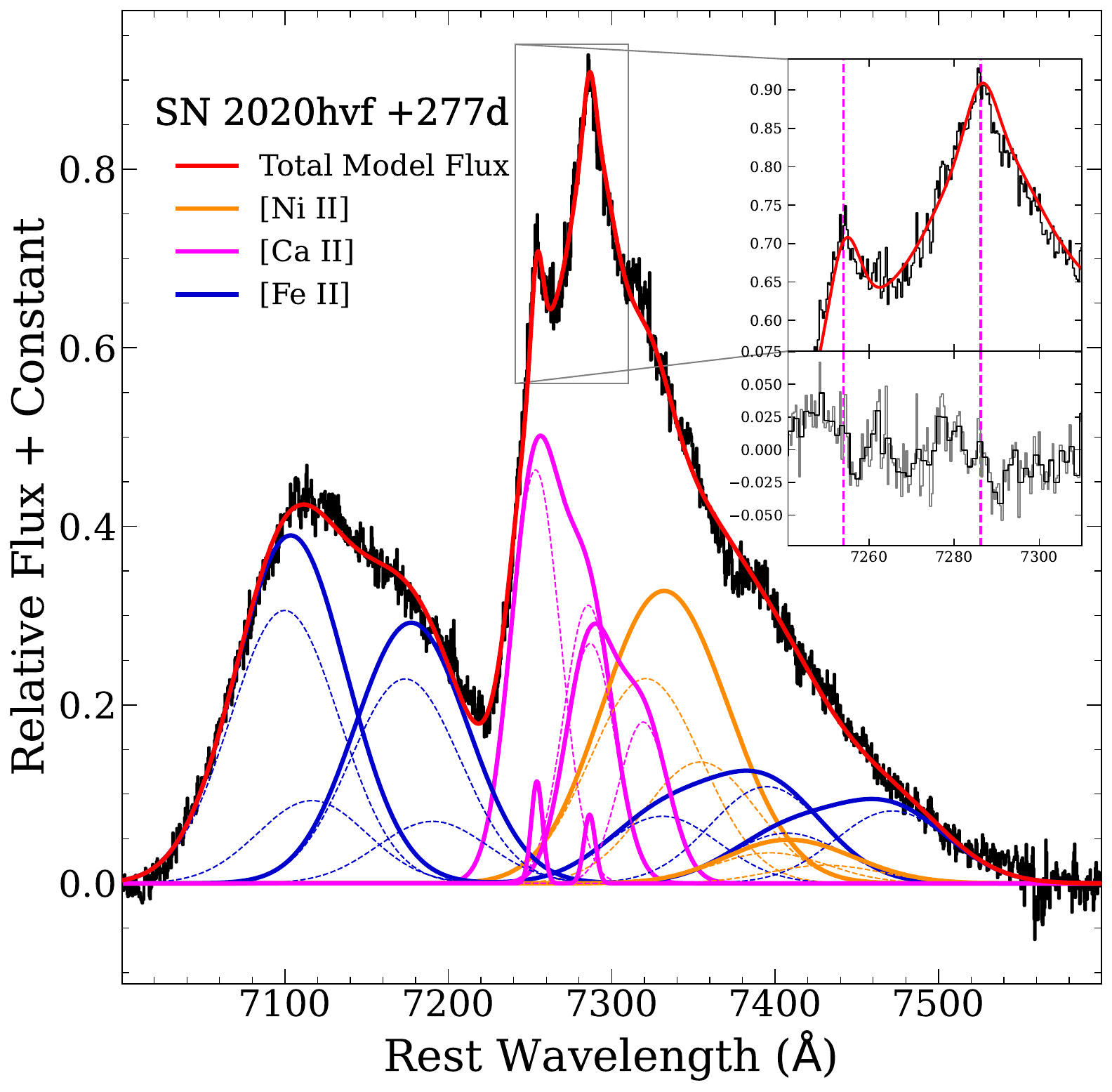}
\caption{Same as \autoref{f:neb_fit}, but the Gaussian fitting method is restricted such that all components of Fe and Ni have the same velocity width, respective blue/red components of Fe and Ni have the same velocity, and both broad components of Ca have the same velocity width. A narrow component of [\ion{Ca}{2}] is still needed to reproduce the observed emission.}\label{f:neb_fit_strict}
\end{center}
\end{figure}

\begin{deluxetable}{lrr}

\tablewidth{0pt}
\tablecaption{Model Parameters for Nebular Emission\label{t:neb_params}}
\tablehead{
\colhead{Parameter} &
\colhead{Two-Component} &
\colhead{Three-Component} \\
\colhead{} &
\colhead{Model} &
\colhead{Model}}

\startdata
Blueshifted [\ion{Fe}{2}]                &                       &                       \\
\hspace{10pt} Offset (km~s$^{-1}$)       & $-2260$ (60)         & $-2230(20)$               \\
\hspace{10pt} Width (km~s$^{-1}$)        & 3390 (40)             & 3400(20)                  \\
\hspace{10pt} Amplitude\tablenotemark{a} & 0.31 (0.01)         & 0.309(0.003)             \\
Redshifted [\ion{Fe}{2}]                 &                       &                       \\
\hspace{10pt} Offset (km~s$^{-1}$)       & 880 (60)               & 840(10)                   \\
\hspace{10pt} Width (km~s$^{-1}$)        & 3390(40)\tablenotemark{b} & 3400(20)\tablenotemark{b} \\
\hspace{10pt} Amplitude                  & 0.22 (0.01)         & 0.216(0.004)           \\
Blueshifted [\ion{Ni}{2}]                &                       &                       \\
\hspace{10pt} Offset (km~s$^{-1}$)       & $-3510(20)$               & $-3510(10)$               \\
\hspace{10pt} Width (km~s$^{-1}$)        & 3100(100)                  & 3080(50)                  \\
\hspace{10pt} Amplitude                  & 0.33(0.01)                  & 0.34(0.01)               \\
Redshifted [\ion{Ni}{2}]                 &                       &                       \\
\hspace{10pt} Offset (km~s$^{-1}$)       & $-1055$               & $-860(10)$               \\
\hspace{10pt} Width (km~s$^{-1}$)        & 4000(200)               & 4000(100)                  \\
\hspace{10pt} Amplitude                  & 0.10(0.01)              & 0.091(0.004)                 \\
Blueshifted [\ion{Ca}{2}]                &                       &                       \\
\hspace{10pt} Offset (km~s$^{-1}$)       & $-1560(20)$               & $-1570(10)$               \\
\hspace{10pt} Width (km~s$^{-1}$)        & 1160(20)                & 1180(20)                   \\
\hspace{10pt} Amplitude                  & 0.44(0.01)               & 0.42(0.01)                  \\
Redshifted [\ion{Ca}{2}]                 &                       &                       \\
\hspace{10pt} Offset (km~s$^{-1}$)       & $-330(30)$                & $-350(20)$                \\
\hspace{10pt} Width (km~s$^{-1}$)        & 800(60)                   & 800(60)                   \\
\hspace{10pt} Amplitude                  & 0.11(0.01)                 & 0.09(0.01)                 \\
Narrow [\ion{Ca}{2}]                     &                       &                       \\
\hspace{10pt} Offset (km~s$^{-1}$)       & \nodata               & $-1560(10)$               \\
\hspace{10pt} Width (km~s$^{-1}$)        & \nodata               & 180(40)                    \\
\hspace{10pt} Amplitude                  & \nodata               & 0.07(0.01)
\enddata

\tablenotetext{a}{Arbitrary units set to match the spectrum as scaled in \autoref{f:neb_fit}.}
\tablenotetext{b}{Width fixed to match the blue-shifted component.}
\end{deluxetable}

In \autoref{f:neb_fit_bad}, we show a rudimentary fit to the 7300\AA\ feature that includes single skewed Gaussian velocity components of [\ion{Fe}{2}] and [\ion{Ni}{2}], and a single Gaussian velocity component of [\ion{Ca}{2}]. Although this model can generally reproduce the line complex, it fails in several ways. First, the bluer portion of the feature, consisting primarily of [\ion{Fe}{2}] emission, has a clearly asymmetric ``saw tooth'' shape where the blue edge is sharper with a skew to the red.  The redder feature also has this general shape, however, that feature is more complex.

While a wide-velocity, skewed line profile for [\ion{Fe}{2}] and [\ion{Ni}{2}] can reproduce the shapes of most of the complex, such a line profile cannot match the [\ion{Ca}{2}] emission. Because the [\ion{Ca}{2}] lines are relatively narrow (and strong), we can clearly see a Gaussian component offset by roughly $-1500$~km~s$^{-1}$. Adding such a component fails to match the line strength of the redder component as well as has a generally poor overall fit to the redder part of the line complex.

Examining the line profile in detail, we notice a shoulder (formally, a slight peak) at $\sim$7320~\AA. By assuming that this shoulder corresponds to [\ion{Ca}{2}] $\lambda$7286 emission, we add a separate [\ion{Ca}{2}] component to our above procedure. Doing so not only well fits the shoulder with $v \approx -300$~km~s$^{-1}$ but at that velocity, the bluer [\ion{Ca}{2}] $\lambda$7254 line overlaps with the [\ion{Ca}{2}] $\lambda$7286 emission from the $-1500$~km~s$^{-1}$ component to produce the correct strength of the emission at 7285~\AA. \hvf{} has two unique [\ion{Ca}{2}] emission components with velocity offsets of roughly $-300$ and $-1500$~km~s$^{-1}$.

Since \hvf{} has separate kinematic components for Ca, it is reasonable to explore if other elements also have separate components. We perform the above procedure again with two components for [\ion{Ca}{2}], [\ion{Fe}{2}], and [\ion{Ni}{2}], but assuming symmetric Gaussian profiles for each species and component. The resulting fit is shown in the left panel of \autoref{f:neb_fit}. We are able to reproduce the broad features of the line complex with such a multi-component model, and this model does not require any skewness for any individual component. While we cannot rule out that the Fe and Ni components are single, skewed components, the Ca has lower velocity offsets consistent with being from the innermost ejecta. It is a reasonable assumption that the Fe and Ni emission are also produced from multiple velocity components.

By zooming in on the [\ion{Ca}{2}] peaks and examining the residuals in \autoref{f:neb_fit}, we find there remain narrow peaks in the emission for which the model does not match. The wavelength offset between these narrow peaks is perfectly matched by [\ion{Ca}{2}], indicating that there exists an additional, narrow ($FWHM < 200$~km~s$^{-1}$) component. Adding such a component to the above model and refitting, we are able to reproduce the entire line profile (right panel, \autoref{f:neb_fit}). The narrow component is found to have a FWHM of $180$~km~s$^{-1}$, a nebular velocity width that is unprecedented in SNe~Ia, but consistent with that of some SNe~Iax \citep{Foley16:iax}. We note that these narrow lines are not from the host galaxy -- they are offset by $-1560$~km~s$^{-1}$ -- and must come from the SN ejecta. An additional fit was performed to investigate the potential for a complementary narrow component associated with the redshifted Ca component, but we found no significant evidence for such a feature. The best fit parameters for each emission model shown in \autoref{f:neb_fit} are compared in Table \ref{t:neb_params}.

While the previous model does an excellent job at reproducing the narrow [\ion{Ca}{2}] features, it does require significant velocity offsets between the [\ion{Fe}{2}] and [\ion{Ni}{2}] distributions ($-2230$~km~s$^{-1}$ and $-3510$~km~s$^{-1}$, respectively for the blue component, and $840$~km~s$^{-1}$ and $-860$~km~s$^{-1}$, respectively for the red component). While similar velocity offsets of $\sim 1000$~km~s$^{-1}$ have been observed in some SNe~Ia \citep{Maguire18}, we also perform a fit with fewer total parameters to see if we can reproduce the feature (\autoref{f:neb_fit_strict}). Here we have forced [\ion{Fe}{2}] and [\ion{Ni}{2}] to have the same velocities and velocity widths. With the restrictions, we find best-fit velocities of $-2330$~km~s$^{-1}$ and $770$~km~s$^{-1}$ for the blue and red components, respectively (similar to those derived for [\ion{Fe}{2}] in our previous fit. We can see that in \autoref{f:neb_fit_strict} with the blue component of [\ion{Ni}{2}] forced to a higher velocity, its amplitude must be decreased, and the velocity width of broad [\ion{Ca}{2}] must be increased to compensate for the lack of flux near 7320\AA. This fit replicates the narrow features in the spectrum well and requires a slightly larger velocity width for the narrow component of [\ion{Ca}{2}] of $340\pm30$~km~s$^{-1}$ (compared with $180\pm40$~km~s$^{-1}$ from the previous fit). However, the red shoulder of the red [\ion{Ca}{2}] component is not well reproduced by this more restrictive fit. Regardless of the complexity of the fit, it is clear that multiple velocity components of each element are required to reproduce the observed 7300\AA\ emission feature.

\begin{figure*}
\begin{center}
    \includegraphics[width=3.2in]{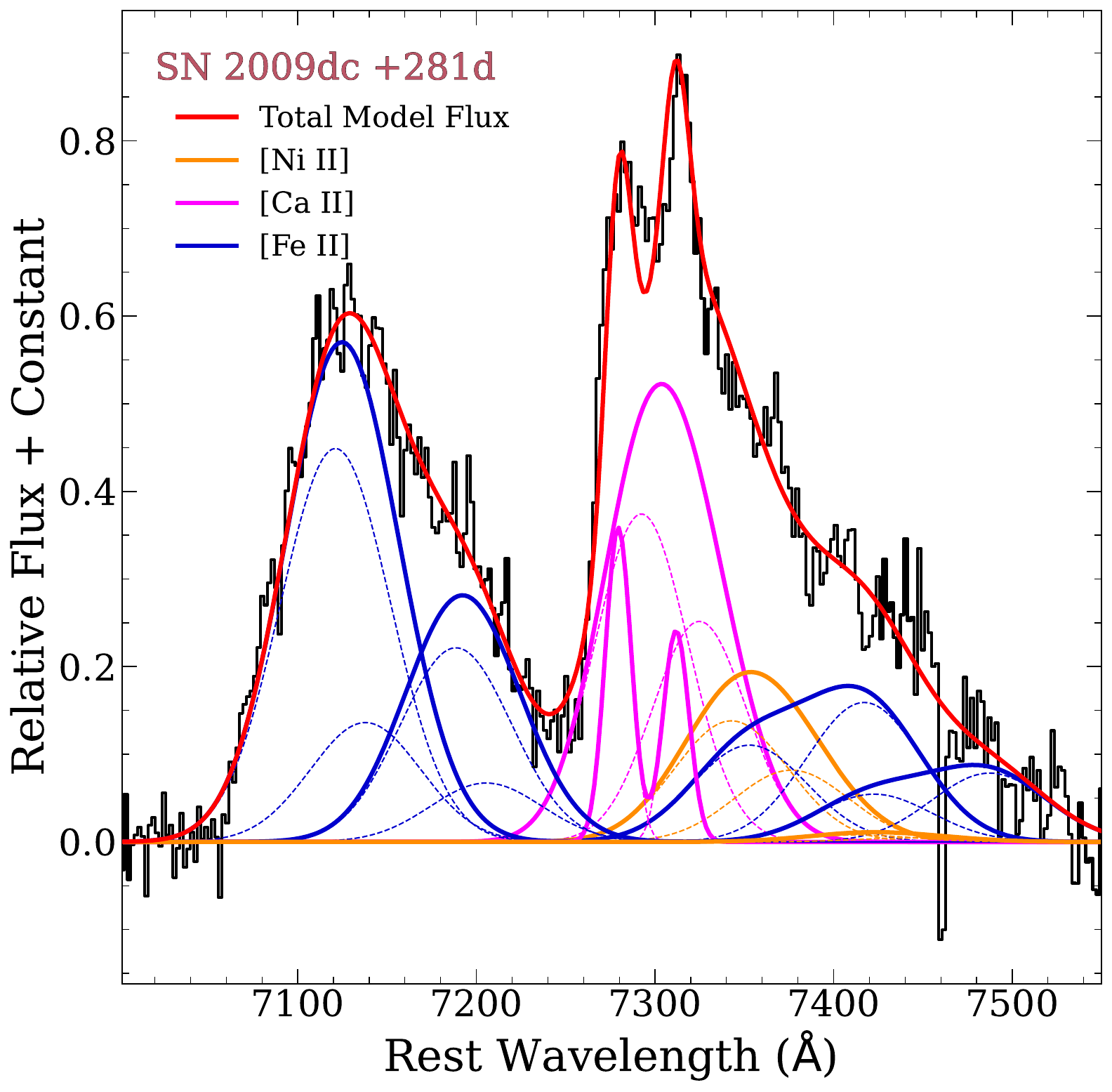}
    \includegraphics[width=3.2in]{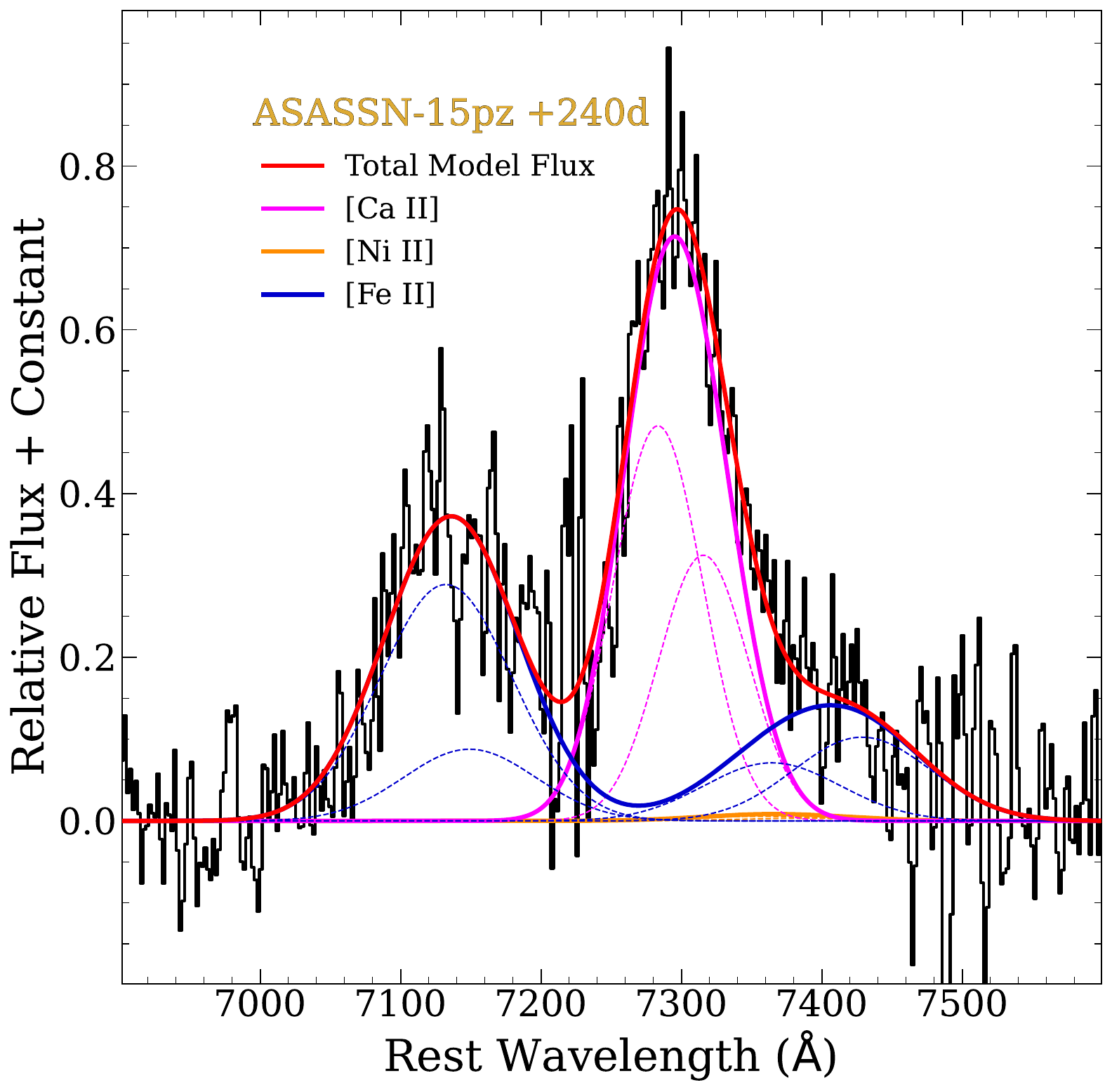}
    \includegraphics[width=3.2in]{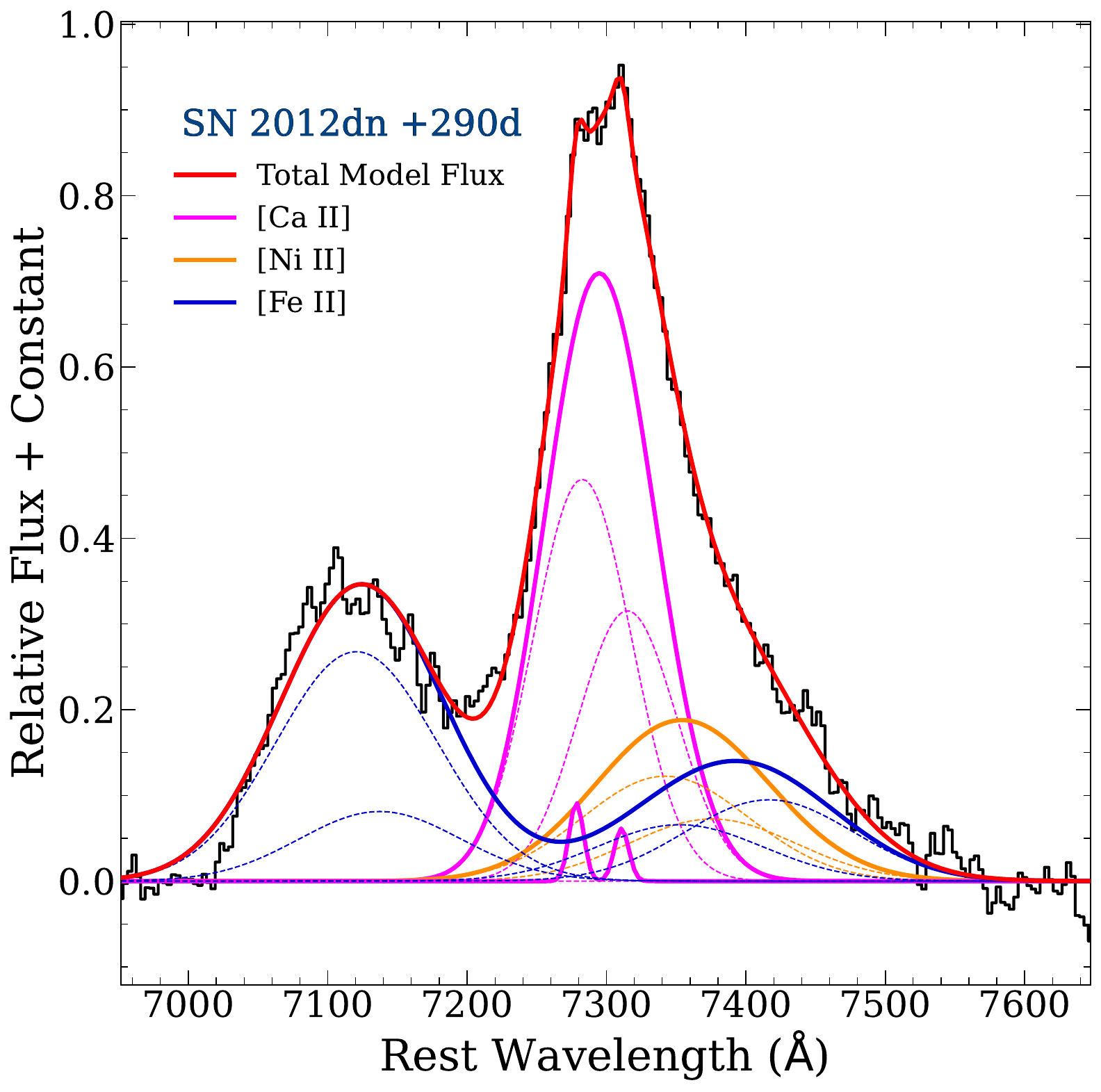}
    \includegraphics[width=3.2in]{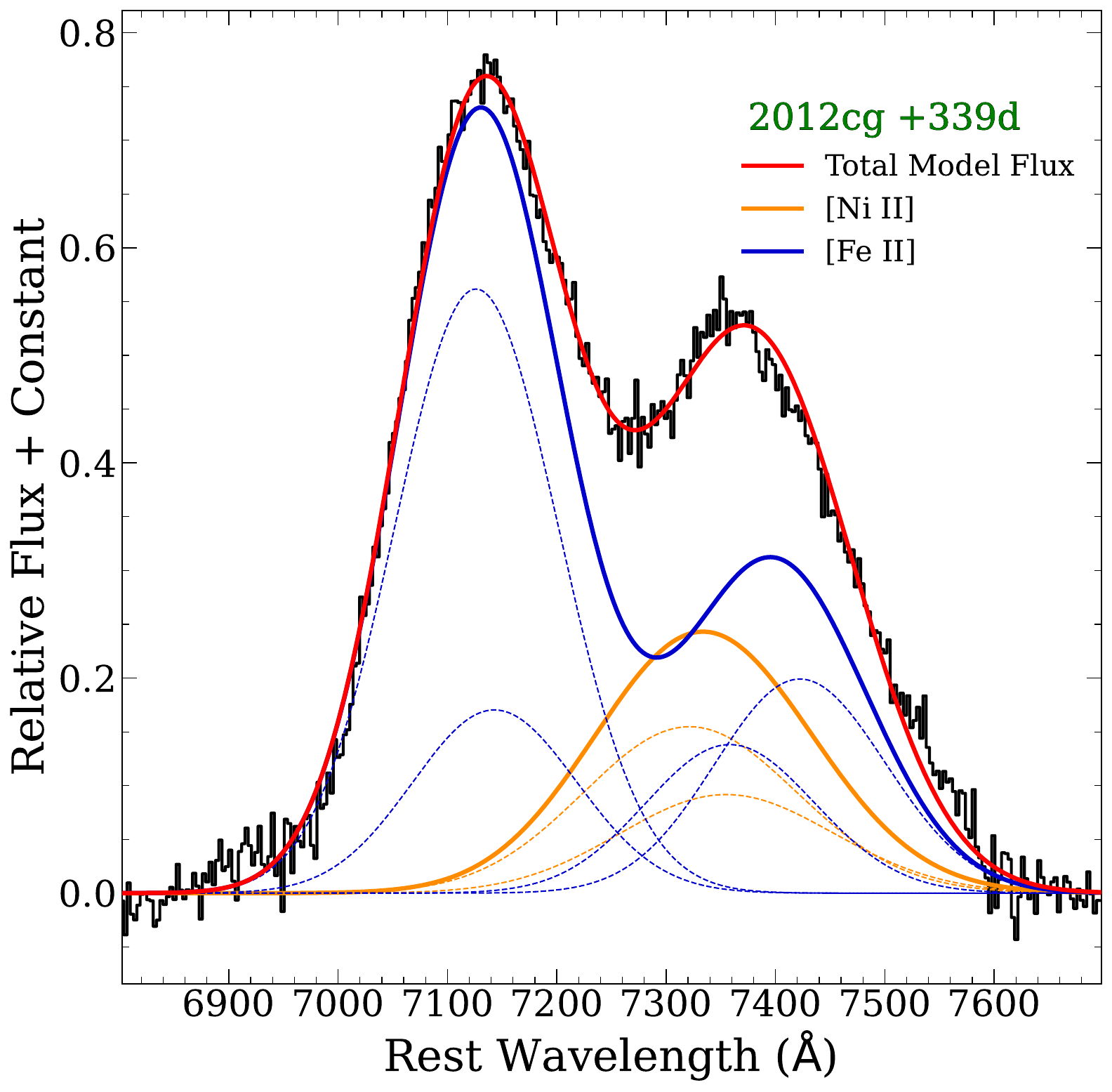}
\caption{Multiple component Gaussian fits to the 7300 \AA\ nebular emission features of SN~2009dc (top left), ASASSN-15pz (top right), SN~2012dn (bottom left), and SN~2012cg (bottom right). While no 03fg-like SNe require extremely narrow velocity components like \hvf{}, multiple components of [\ion{Ca}{2}] are required to reproduced the emission observed in SN~2009dc, and SN~2012dn.}\label{f:neb_fit_others}
\end{center}
\vspace{-1mm}
\end{figure*}


A similar ``saw tooth'' profile is seen in other 03fg-like~SNe~Ia such as SNe~2009dc and 2012dn. To determine if these objects also had multiple velocity components, we performed the same fitting procedure as above for their corresponding nebular spectra. These fits are displayed in the top-left and bottom-left panels of \autoref{f:neb_fit_others}, respectively. Here we also include additional nebular line fits to ASASSN-15pz (top-right), another 03fg-like SN~Ia, and SN~2012cg (bottom-right), a high-luminosity SN~Ia with an early flux excess and nebular features characteristic of normal-SNe~Ia. While we can fit these spectra with distinct velocity components, this is not always required by the data. In particular, the [\ion{Fe}{2}] feature in SN~2009dc, shares a similar skewness to \hvf{}, and when fit with two velocity components, yields blue- and redshifted velocity offsets of $-1430 \pm 60$~km~s$^{-1}$ and $1400 \pm 30$~km~s$^{-1}$, respectively. Additionally, the full peak near $\sim 7320$\AA\ is difficult to explain without including both a broad and narrow component of [\ion{Ca}{2}]. In this case, the width of the best-fit narrow component is $700 \pm 20$~km~s$^{-1}$ which is significantly larger than what we found for \hvf{}. Similarly, for SN~2012dn, we find that including both broad and narrow components of [\ion{Ca}{2}] is necessary to reproduce the peak near $\sim 7320$\AA\ and we determine that the width of the best-fit narrow component is $500$~km~s$^{-1}$. Similar narrow emission in the nebular phase has also been observed in the Ca-strong sub-luminous SN~2016hnk \citep{Galbany19}. At +351 days, the authors measure [\ion{Ca}{2}] with a width of $700$~km~s$^{-1}$ and attribute this emission to the presence of Ca$^{48}$ within the central regions of the ejecta. SN~2012dn has less evidence for multiple [\ion{Fe}{2}] or [\ion{Ni}{2}] peaks and these species can be fit by single components. Nevertheless, the saw tooth profile in these SNe is indicative of some asymmetric emission. For ASASSN-15pz, we require a strong component of [\ion{Ca}{2}] (reminiscent of the 03fg-like SN Ia 2020esm, \citealt{Dimitriadis21}). If we instead attempt to fit the feature near 7320\AA\ with [\ion{Ni}{2}], we find [\ion{Fe}{2}] and [\ion{Ni}{2}] velocities of $-960$~km~s$^{-1}$ and $-3800$~km~s$^{-1}$. This velocity difference would be one of the largest measured in a SN~Ia \citep{Maguire18}. For comparison, SN~2012cg displays a similar simple 7300\AA\ emission feature and has best fit [\ion{Fe}{2}] and [\ion{Ni}{2}] velocities of $-1220$~km~s$^{-1}$ and $-2350$~km~s$^{-1}$. Given that 03fg-like SNe tend to have [\ion{Ca}{2}] emission in their nebular spectra, we expect that this feature is more likely dominated by [\ion{Ca}{2}] rather than [\ion{Ni}{2}]. Additionally in ASASSN-15pz, we do not detect any evidence of asymmetry, however, we note that at earlier phases, ASASSN-15pz displayed extremely narrow [\ion{Ca}{2}] and Ca II NIR triplet emission (both $\sim200$ ~km~s$^{-1}$, \citealt{Chen19}). Also, in all cases where a broad component of [\ion{Ca}{2}] is required to fit this feature, its best-fit velocity width is smaller than the corresponding [\ion{Fe}{2}] and [\ion{Ni}{2}] components. This was also true for SN~2019yvq \citep{Siebert20b}. This could indicate that the ejecta is stratified with Calcium only appearing closer to the center. Future studies examining a large sample of 03fg-like SNe~Ia should both carefully look for multiple components of these species, and further investigate if the saw-tooth pattern where the emission is skewed red is ubiquitous.

\subsection{Spectroscopic Evolution}

Our observations are able to constrain the evolution of the nebular spectroscopic properties of \hvf{} over the course of $\sim100$ days. The full optical spectra are presented in \autoref{f:neb_spec}. The most striking change from $+240$ to $332$ days after maximum light is the evolution of the [\ion{Ca}{2}]/[\ion{Fe}{3}] ratio. Over 31 days this ratio (measured from the peaks of the 7300\AA\ and 4700\AA\ features) changes from 0.40 to 0.82 and then remains constant until our last spectrum 61 days later. We attribute this change to the changing ionization state of Fe as the ejecta expands and cools. However, given that the ratio of these lines has been used to infer progenitor scenarios \citep{Polin21:neb}, we emphasize that sequences of spectra should be obtained in the nebular phase to measure potential evolution.

\begin{figure}
\begin{center}
    \includegraphics[width=3.2in]{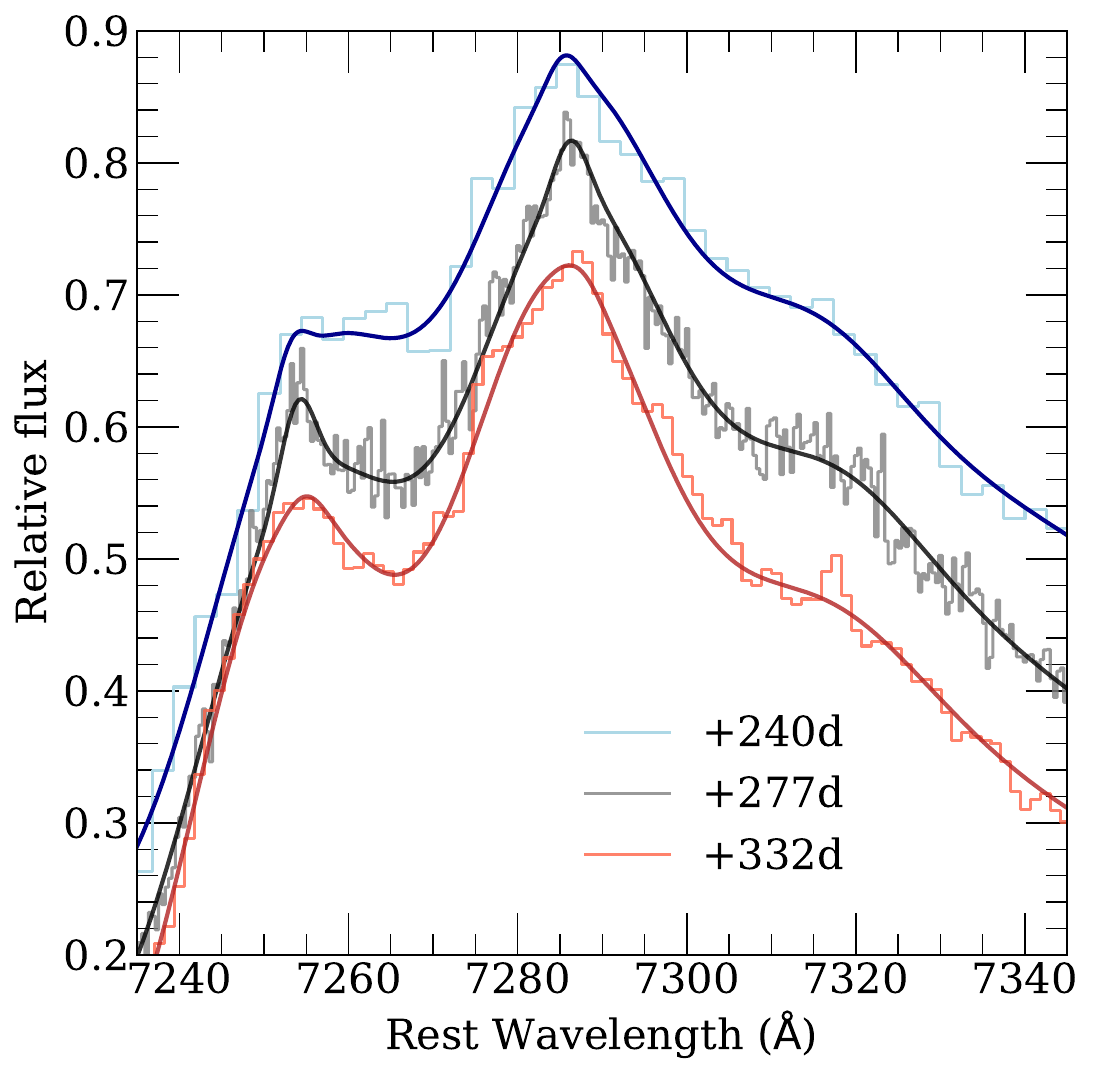}
\caption{Nebular phase spectra of \hvf{} ranging from +240-332 days zoomed in on the [\ion{Ca}{2}] emission region. The light blue, gray, and light red curves are spectra obtained from the Kast spectrograph ($R\sim3000$), the higher resolution grating on the LRIS spectrograph ($R\sim12000$), and a low-resolution grating on the LRIS spectrograph ($R\sim6000$). While narrow [\ion{Ca}{2}] may be present in each of these spectra, we are only able to constrain it with the +277d higher resolution spectrum.}\label{f:neb_fit_time}
\end{center}
\end{figure}

Our sequence of nebular spectra also provides an opportunity to measure complex [\ion{Ca}{2}] emissions from multiple sources. In \autoref{f:neb_fit_time} we display spectra from three different sources (Kast, 300/7500 grating; light blue curve; LRIS, R1200/7500 grating, gray curve; and LRIS R400/8500, light red curve) zoomed in on the main peak of the 7300\AA\ emission feature. The varying spectral resolution of these observations limits our ability to constrain the narrow [\ion{Ca}{2}] component. Using the sky lines present in the data, we measure a Gaussian line spread function for each instrumental setup. We find that these line spread functions have FWHM widths of 380~km~s$^{-1}$, 240~km~s$^{-1}$, and 70~km~s$^{-1}$, for Kast 300/7500, LRIS R400/8500, and LRIS R1200/7500, respectively. When the same 3-Ca component fit is applied to each of these observations (accounting for these line-spread functions), we find that only the highest resolution LRIS spectrum is able to constrain the narrow component of [\ion{Ca}{2}] (180~km~s$^{-1}$). The broad features of Fe, Ni, and Ca measured from these fits, do not show any significant evolution with time.

\section{Discussion}\label{s:disc}

\subsection{A Basic Progenitor and Explosion Model}

The detection of multiple velocity components in the nebular spectra of \hvf{} along with the identification of a low-velocity width [\ion{Ca}{2}] feature provides strong constraints on the progenitor system and SN explosion. At a basic level, the two main velocity components with offset velocities of $-1570$ and $-350$~km~s$^{-1}$ relative to the host-galaxy systemic velocity, respectively, is a direct confirmation of an asymmetric explosion. This fact on its own points toward a DD progenitor system, and given the total ejecta mass of 2.1 M$_{\odot}$ estimated by \citet{Jiang+21_SN20hvf}, requires large WDs (both $>1.0$ $M_{\odot}$ assuming equal masses). We note that this ejecta mass is primarily estimated from the peak luminosity and light-curve shape. Significant luminosity produced through the interaction of the ejecta with the CSM may allow for a smaller total ejecta mass, and thus the smaller WDs. The narrow [\ion{Ca}{2}] component with ${\rm FWHM} \approx 180$~km~s$^{-1}$ indicates the presence of ejecta with very low kinetic energy. Below we examine in more detail the implications of these observations.

The narrow [\ion{Ca}{2}] emission that represents low-energy ejecta is offset in velocity by $-1560$~km~s$^{-1}$, coincident with the blueshifted broad velocity features. Combined with its low width, this line must originate from an ejecta component that is moving quickly towards us yet with minimal velocity spread. Such a configuration can only be achieved if the center of this component is offset from the center of mass of the full explosion and, relative to that offset, the ejecta has a very low velocity.


The small velocity width for the narrow [\ion{Ca}{2}] feature requires barely unbound ejecta. While it is possible that both progenitor stars were completely disrupted but with (at least) one barely unbound, this is unlikely. This would both require a careful balance of the explosion energy and gravitational potential energy and is inconsistent with the energy density of C/O nuclear burning. The most likely scenario is that one WD remained bound and this component represents a wind from the surface of the WD. In such a scenario, the surviving WD is likely ``puffed up" with a radius significantly larger than its pre-explosion radius. Thus, the physics of this surviving star is directly comparable to the bound remnants that have been inferred from the late time observations of SNe~Iax \citep{Foley13:iax, Foley14:08ha, Foley16:iax}. The 3D simulations in \citet{Fink14} show that pure deflagrations of WDs produce bound remnants that can range in size from 0.102 - 1.32 $M_{\sun}$, and observational evidence suggests the radius could range from anywhere between 1 $R_{\sun}$ \citep{Maeda22} and $10^4$ $R_{\sun}$ \citep{Foley16:iax}. While the size of the remnant is highly uncertain, we can place additional constraints on \hvf{} using the limit on the total ejecta mass of 2.1 $M_{\sun}$ \citep{Jiang+21_SN20hvf}, and the knowledge that the progenitor system is likely DD. We can rule out the maximum bound remnant mass, 1.32 $M_{\sun}$, since this scenario would indicate a total mass of $2.1 + 1.32 = 3.42~{\rm M}_{\odot}$, requiring an individual WD mass $>1.71$ $M_{\sun}$. We determine an upper limit on the total mass by requiring the maximum individual WD mass to be the maximum mass of a CO white dwarf (1.21 $M_{\sun}$, \citealt{Nelemans01}). This provides an upper limit on the bound remnant mass of 0.32 $M_{\sun}$. Thus if we assume the mass in the middle of our allowed range (0.1 - 0.3 $M_{\sun}$) we find a radius for the bound remnant that is consistent with SN Iax observations:
\begin{linenomath*}
\begin{equation}
  R_{b} = 9 R_{\sun} \, \left ( \frac{v_{\rm esc}}{90 \rm{~km~s}^{-1}} \right )^{-2} \left ( \frac{M}{0.2 M_{\sun}} \right )
\end{equation}
\end{linenomath*}
where $v_{\rm esc}$ and $M$ are the escape velocity of the ejecta (estimated from the HWHM of the narrow Ca component) and the mass of the surviving star, respectively.

We take the center-of-mass LOS velocity of the system to be the average of the offset velocities of the blue and redshifted broad components of [\ion{Ca}{2}]. We measure the difference between these velocity components to be $1220$~km~s$^{-1}$ corresponding to the LOS difference in orbital velocity ($v_{\rm orb}$) of the two components. Therefore, we define
\begin{linenomath*}
\begin{equation}
    v_{\rm orb} \sin i \cos \theta = v_{\rm meas}  = 610 {\rm ~km~s}^{-1},
\end{equation}
\end{linenomath*}
where $i$ is the orbital inclination and $\theta$ is the angle between our line of sight and the velocity vector in the orbital plane. Since the broad blueshifted ejecta component has a similar offset velocity to that of the narrow component, the material associated with the broad blueshifted emission is most likely material originating from the WD that produced a bound remnant \citep{Fink14,Pakmor_Zenati+21, Burmester+23}. In this case, there would be a surface explosion that does not completely unbind the star but does produce a significant amount of ejecta at a similar systemic velocity. 

\begin{figure}
\begin{center}
    \includegraphics[width=3.2in]{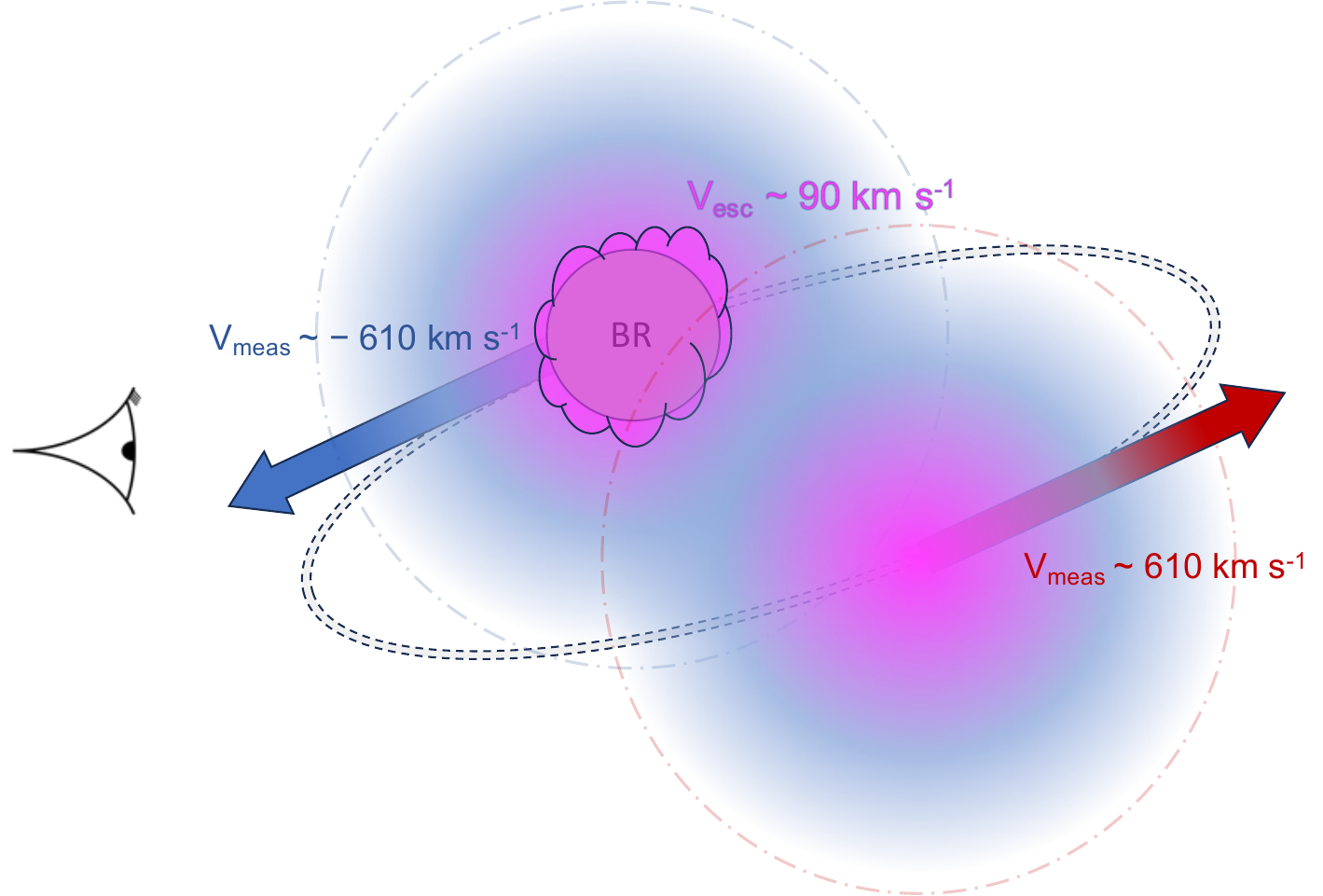}
\caption{Schematic diagram (not to scale) for the nebular phase ejecta distribution of \hvf{}. Pink and blue shaded regions correspond to [\ion{Ca}{2}] and [\ion{Fe}{2}] + [\ion{Ni}{2}] line emitting regions, respecitively. Blue and red dashed lines indicate the extent of the blue and redshifted ejecta. A bound remnant (BR) with a Ca wind ($v_{\rm esc} = 90$ \kms) is depicted at the center of the blueshifted component. The black-dashed oval indicates the initial orbital path of the progenitor system.}\label{f:schem}
\end{center}
\end{figure}

In this scenario, the redshifted component would then represent material from the second WD. This interpretation requires an orientation with a significant orbital velocity component along our LOS (i.e., large $i$ and small $\theta$). This scenario is illustrated in more detail in \autoref{f:schem} where we present a schematic diagram that highlights the specific unique properties of our inferred ejecta distribution. If the asymmetry was instead in the plane of the sky ($i \approx 0$ or $\theta \approx \pi/2$), we would expect smaller velocity offsets that would be more difficult to detect. While \hvf, SN~2009dc, and SN~2012dn all require asymmetric systems to explain their emission, the nebular emission seen in the 03fg-like SNe ASASSN-15pz (\autoref{f:neb_fit_others}), SN~2020esm \citep{Dimitriadis21}, and SN~2021zny \citep{Dimitriadis23} seem to be well-explained with single velocity components, potentially a result of this orientation effect. 

As we do not detect any narrow redshifted emission features, the star associated with that component was likely fully disrupted. It may be surprising that the inferred surviving companion is associated with the stronger broad component of [\ion{Ca}{2}] since, nominally, this would imply more radioactive material was synthesized at that location. However, some have suggested that ``saw-tooth" emission profiles seen in some 03fg-like SNe could be the result of dust obscuring the far side of the ejecta resulting in fainter redshifted features \citep{Taubenberger13,Taubenberger19}. If bound remnants are common for 03fg-like SNe, some SNe should have narrow components associated with redshifted components. 

With the orbital velocity, we can estimate the separation between the two WDs at the time of explosion, $a$, using
\begin{linenomath*}
\begin{equation}
  v_{\rm orb} = \sqrt{\frac{Gm_{1}^2}{(m_{1}+m_{2})a}},\label{e:vorb}
\end{equation}
\end{linenomath*}
where $m_1$ and $m_2$ are the individual WD masses. While we cannot constrain the mass ratio of the system, \citet{Jiang+21_SN20hvf} estimated that the total ejecta mass, $M_{\rm ej}$, was 2.1~$M_{\odot}$ and we have find that the mass left in a bound remnant to be $M_{b} \approx 0.2 M_{\odot}$. Therefore,
\begin{linenomath*}
\begin{equation}
    m_1 + m_2 = M_{\rm ej} + M_{b} = 2.3~{\rm M}_{\odot}.
\end{equation}
\end{linenomath*}

Assuming an equal mass system, we can rewrite Equation~\ref{e:vorb} in terms of only measured properties,
\begin{linenomath*}
\begin{align}
  a &= \frac{G(M_{\rm ej}+ M_{b}) \sin^2 i \cos^2 \theta}{4 v_{\rm meas}^2}\\
  &= 0.6 \times 10^{9} {\rm ~cm~} \frac{M_{\rm tot}}{2.3 {\rm ~M}_{\odot}} \left ( \frac{v_{\rm meas}}{610 {\rm ~km~s}^{-1}} \right )^{-2} \notag \\
  & {~} \times \frac{\sin^2 i}{\sin^2 \pi/4 } \frac{\cos^2 \theta}{\cos^2 \pi/4 }. \label{e:sep}
\end{align}
\end{linenomath*}
where we have scaled the value to a typical inclination angle of $i=\pi/4$ and a LOS angle of $\theta=\pi/4$. We emphasize again that a smaller total mass is possible if some of the peak luminosity of the SN can be attributed to CSM interaction. This would lead to a lower estimate for the binary separation.

We can also estimate what the binary separation should be during Roche-Lobe overflow as approximated by \citet{Eggleton83}. If here we also assume equal masses, we find $a = 1.4\times 10^9$~cm, consistent with the above result. Our estimate of the separation is also consistent with simulations of mass transfer in DD systems. For example, \citet{Marius11} found initial separations required for initiating mass transfer ranging from $1.91-5.77\times 10^9$ cm with their highest mass system ($M_{\rm tot} = 2.1$ with a 0.9 and 1.2~M$_{\odot}$ C/O WDs) having the smallest separation.

Under the same assumptions as above, we can determine the orbital period at the time of the explosion,

\begin{linenomath*}
\begin{align}
  t &= 2.1 {\rm ~min} \, \frac{m_1}{1.15 {\rm ~M}_{\sun}} \left ( \frac{v_{\rm meas}}{610 {\rm ~km~s}^{-1}} \right )^{-3} \notag \\
  &{~} \times \frac{\sin^3 i }{\sin^3 \pi/4} \frac{\cos^3 \theta }{\cos^3 \pi/4}.
\end{align}
\end{linenomath*}

Although this value is highly dependent on the exact measured velocity and orbital inclination, our simple assumptions result in a period that is similar to minimum period observed in AM CVn stars \citep[$\sim$5~min;][]{Ramsay18}. The kinematics measured from the nebular spectrum and some simple assumptions about the progenitor system naturally point toward dynamical properties that are consistent with those predicted from binary WD progenitors. 

Several of the inferred properties of \hvf{} allow us to narrow down potential explosion scenarios. The large ejecta mass \citep{Jiang+21_SN20hvf} likely requires a DD progenitor. The high luminosity ($M_B = -19.9$~mag) requires a large amount of synthesized $^{56}$Ni. The early bump in the light curve and unburned material seen in early optical spectra indicate the presence of dense CSM \citep{Jiang+21_SN20hvf}. The presence of strong [\ion{Ca}{2}] emission and weak [\ion{Fe}{3}] emission at late times likely indicate high-density burning as seen in SN~2009dc \citep{Taubenberger13}. Furthermore, the multiple broad components of [\ion{Ca}{2}] indicate that significant burning occurred at locations of both the primary and secondary WD. Finally, the narrow component of [\ion{Ca}{2}] points toward an explosion model that must be capable of producing a surviving companion. 

One potential DD explosion model is the ``double-detonation" scenario where a surface He-shell detonation triggers a carbon detonation in the core of a sub-Chandrasekhar mass WD \citep{Guillochon10, Woosley11, MariusDan+11, DanM+12, Nomoto18, Polin19, Perets_Zenati+19}. \citet{Shen18a}, showed that the detonation of sub-M$_{\rm Ch}$ mass bare C/O WDs can explain the observed optical properties of normal SNe~Ia. Multiple studies have looked for surviving companion stars produced by this scenario \citep{Shen18b, ElBadry+23_HVS}. \citet{ElBadry+23_HVS} found that DD binaries with total mass in excess of the M$_{\rm Ch}$ are the dominant fraction of observed hyper-velocity WDs in the MW. They also suggest that multiple progenitor channels may be needed to explain the total population. While \hvf\ likely has a surviving companion WD, it is unlikely that this model is capable of producing both the total ejecta mass and early time flux excess. Additionally, while we infer a smaller velocity than expected for runaway stars from higher mass systems, we note that our measurement of $610$~km~s$^{-1}$ is a minimum on the orbital velocity whose exact value depends on the inclination angle. This minimum does, however, effectively rule out main sequence or core He-burning stars as donors \citep{Shen18b}, providing additional strong evidence for a DD progenitor system.

Similar modeling has been performed to understand what happens if the primary WD has a more massive He shell \citep{Polin19, Polin21:neb}. This double-detonation model is expected to produce early excess flux due to the presence of radioactive material in the ashes of the He shell. \citet{Polin21:neb} find that the mass of the He shell can determine the amount of expected nebular [\ion{Ca}{2}] emission. \citet{Siebert20b} found that this model could explain the nebular spectrum of SN~2019yvq, however no model could adequately explain all of its observed properties \citep{Miller20,Tucker21,Burke21}. While there are similarities between SN~2019yvq and \hvf{}, a sub-M$_{\rm Ch}$ double-detonation with a peak luminosity comparable to \hvf{} is expected to have a very high \ion{Si}{2} velocity at peak brightness ($\sim$16,000~km~s$^{-1}$), which is inconsistent with the velocity measured for \hvf{}. 

Another potential DD progenitor system is the disruption of a massive hybrid He-C/O WD by a C/O WD companion \citep{Perets_Zenati+19}. \citet{Pakmor_Zenati+21} used a 3D simulation to study the interaction of a He-rich hybrid 0.69 $M_{\odot}$ He-C/O WD with a more massive 0.8 $M_{\odot}$ C/O WD. In this simulation, the interaction leads to a He detonation in the accretion stream which compresses the He-C/O WD resulting in a second detonation that fully disrupts the WD \citep{Guillochon10} leaving the primary WD intact with $M_{b} \approx 0.6$~M$_\odot$ and velocity of $\sim$1000~km~s$^{-1}$. In this scenario, detonations occur for both WDs, potentially leading to multiple components of SN ejecta that are offset by their orbital velocities. This simulation produced a faint and reddened transient inconsistent with the properties of \hvf, therefore we encourage further modeling of this scenario to understand if higher mass systems can explain both the high-luminosity and nebular emission seen in \hvf. 

\vspace{-2mm}
\subsection{Progenitors of 03fg-like SNe~Ia}

As shown in \autoref{f:spec_comp_7300}, other 03fg-like SNe~Ia show strange Ca features in their nebular spectra. In particular, SNe~2009dc and 2012dn require multiple velocity components of [\ion{Ca}{2}] to reproduce the observed emission, while ASASSN-15pz does not. \citet{Jiang+21_SN20hvf} suggest that \hvf\ produced 2.1 $M_{\odot}$ of ejecta and was likely the result of the explosion of a $>$M$_{\rm Ch}$ DD WD system
(similar to what was concluded for SN~2009dc; \citealt{Taubenberger13}). If instead \hvf\ was the result of a sub-M$_{\rm Ch}$ mass explosion in a DD system, it would require large C/O WD primary and secondary masses (likely both with $M > 1.0$~M$_{\odot}$) to be capable of producing the inferred ejecta mass. We estimate a low (albeit uncertain) mass for the surviving companion star ($0.2 M_{\sun}$) and thus cannot rule out a sub-M$_{ch}$ mass origin. 

\citet{Srivastav23} examined the early flux excess in the light curve of SN~2022ilv in detail and came to a similar conclusion for that event. Despite having very different photospheric velocities, \hvf{} and SN~2022ilv have very similar bolometric light curves and similar fast-evolving early excesses \citep{Srivastav23}. \citet{Raskin13} showed that in DD merger simulations, tidal stripping of the secondary WD can result in a C/O-rich CSM nearby to the progenitor system. Interaction with this kind of CSM can explain both the early flux excess and high luminosity of 03fg-like SNe~Ia \citep{Dimitriadis21, Dimitriadis23}. For both, \hvf{} and SN~2022ilv, these previous studies estimated a CSM distance of $10^{13}$ cm from the early bump. This progenitor model requires the complete disruption of the secondary star. Given that the estimated CSM distance is far enough away to require a delay between the disruption and explosion that is larger than the dynamical time, it is hard to reconcile with the existence of a bound remnant that we infer for \hvf.

Alternatively, \hvf{} could be the result of a normal-M$_{\rm Ch}$ mass explosion within a C/O-rich envelope. The lack of \ion{C}{2} absorption at maximum light in combination with its large \ion{Si}{2} velocity \citep{Jiang+21_SN20hvf} is consistent with the correlation between these properties observed for a sample of 03fg-like SNe~Ia \citep{Ashall21}. However, we observe no signs of interaction (e.g., narrow H/He features) of the ejecta with the circumstellar envelope, and the ejecta kinematics suggest a large asymmetry that cannot be explained by this progenitor scenario.

\section{Conclusions}\label{s:conc}
Our nebular spectra of \hvf{} add to the growing evidence that the rare 03fg-like SNe~Ia are a diverse subclass. Thorough investigation of the ejecta kinematics at late times reveal that this event was likely the result of a $M_{\rm tot} > 1.4$~M$_{\odot}$ DD progenitor system that produced a bound remnant. We summarize our analysis of the \hvf{} nebular spectra below:

\begin{itemize}
    \item The nebular emission features of \hvf{} are qualitatively similar to those of other 03fg-like and 02es-like SNe~Ia. Both of these subclasses tend to have high [\ion{Ca}{2}]/[\ion{Fe}{3}] ratios and often display strong emission from both [\ion{Ca}{2}] and the Ca II NIR triplet.  These features require an explosion that differs from typical SN~Ia explosions.  In particular, significant explosive He burning may be required to produce such strong [\ion{Ca}{2}] emission.
    \item The 7300~\AA\ emission feature can only be explained by the presence of multiple velocity components of [\ion{Ca}{2}], requiring multiple, kinematically distinct, ejecta components that are likely the result of an asymmetric explosion.
    \item A low-velocity component (${\rm FWHM} = 180\pm40$~km~s$^{-1}$) is required to explain the sharpest [\ion{Ca}{2}] emission peaks in our spectra. This component could only be resolved with the higher-resolution LRIS spectrum. This extremely narrow emission is similar to that seen in SNe~Iax, and the low velocity is best explained by being roughly the escape velocity of a bound remnant with $M \approx 0.2$~M$_{\odot}$ and $R \approx 10$~R$_{\odot}$. 
    \item We potentially identify weak [\ion{O}{1}] emission. Currently, [\ion{O}{1}] emission has been observed in two 02es-like SNe (SN~2010lp and iPTF14atg), and two 03fg-like SNe (SNe~2012dn, and 2021zny).  The emission is narrow, and offset at a velocity similar to that of the [\ion{Ca}{2}] emission, and thus the oxygen may be unburned material from the bound remnant.
    \item We do not detect any H/He emission indicating that the early excess flux seen for \hvf{} was not caused by interaction with a H- or He-rich CSM, further favoring a 
    DD progenitor system for \hvf.
\end{itemize}

These results emphasize the necessity for nebular spectroscopy in addition to high-cadence early photometry and spectroscopy in order to infer progenitor properties.  03fg-like and 02es-like SNe~Ia have significantly different peak luminosities and likely require different progenitor and/or explosion models, despite their similar early flux excesses. Currently, companion interaction, circumstellar material (CSM) interaction, surface Ni, and double-detonation models all have the capability of producing early flux excesses, but the large variety of early flux excess morphologies \citep{Jiang+18} make inferring progenitor systems with only photometric information very difficult. 

Nebular-phase observations, when the ejecta is optically thin, are particularly determinative for asymmetric velocity distributions, ejecta ionization states, and companion/CSM interaction. For 03fg-like SN~Ia specifically, higher-resolution spectra than is typically obtained are critical to detect potential low-velocity ejecta components like the one seen in \hvf. Expanding our wavelength coverage at late times will also be important for this subclass of SNe~Ia. Nebular NIR and MIR spectra provide unique constraints on velocity distributions that cannot be obtained from optical spectra \citep{Maguire18}. Some features at these longer wavelengths are isolated and thus easier to model \citep{Kwok+23K, Derkacy+23}. Furthermore, \citet{Derkacy+23} found that the velocity extent of stable Ni features in {\it JWST} MIR spectra of the normal SN~Ia~2021aefx are consistent with high-density burning, indicating a near-$M_{\rm Ch}$ mass WD. A MIR spectrum of a likely super-$M_{\rm Ch}$ SN~Ia, would provide strong constraints on the progenitors of both classes of SNe~Ia, and additionally, determine if these SN~Ia form dust at late times.

\acknowledgments
M.R.S.\ is supported by the STScI Postdoctoral Fellowship. Y.Z.\ partially supported by NASA grant NNH17ZDA001N.  G.D.\ is supported by the H2020 European Research Council grant no.\ 758638.
The UCSC team is supported in part by NASA grant NNG-17PX03C, NSF grant AST-1815935, the Gordon and Betty Moore Foundation, the Heising-Simons Foundation, and by a fellowship from the David and Lucile Packard Foundation to R.J.F.

The Computational HEP program in The Department of Energy's Science Office of High Energy Physics provided resources through Grant $\rm \#KA2401022$. Calculations presented in this paper used resources of the National Energy Research Scientific Computing Center (NERSC), which is supported by the Office of Science of the U.S. Department of Energy under Contract No.\ DE-AC02-05CH11231.

The data presented herein were obtained at the W.~M.\ Keck Observatory, which is operated as a scientific partnership among the California Institute of Technology, the University of California and the National Aeronautics and Space Administration. The Observatory was made possible by the generous financial support of the W.~M.\ Keck Foundation. The authors wish to recognize and acknowledge the very significant cultural role and reverence that the summit of Maunakea has always had within the indigenous Hawaiian community. We are most fortunate to have the opportunity to conduct observations from this mountain.

A major upgrade of the Kast spectrograph on the Shane 3~m telescope at Lick Observatory was made possible through generous gifts from the Heising-Simons Foundation as well as William and Marina Kast. Research at Lick Observatory is partially supported by a generous gift from Google.

\facilities{Keck:I (LRIS), Shane(Kast)}
\software{astropy \citep{astropy}}
\newpage
\bibliography{astro_refs}
\bibliographystyle{aasjournal}

\end{document}